\documentclass[prc,twocolumn,showpacs,superscriptaddress]{revtex4-2}
\usepackage{dsfont}
\usepackage{graphicx,amsmath}
\def\be{\begin{eqnarray}}
\def\ee{\end{eqnarray}}
\def\bc{\begin{center}}
\def\ec{\end{center}}

\newcommand{\lsim}{\stackrel{\scriptstyle <}{\phantom{}_{\sim}}}
\newcommand{\gsim}{\stackrel{\scriptstyle >}{\phantom{}_{\sim}}}

\usepackage{xcolor}

\begin{document}
\title{Pion-sigma meson vortices in rotating  systems}
\author{D. N. Voskresensky}
   \affiliation{ Joint Institute for Nuclear Research, Joliot-Curie street 6,
		141980 Dubna, Russia}
 \affiliation{ National Research Nuclear
    University (MEPhI),  Kashirskoe shosse 31,
    115409 Moscow, Russia}

\begin{abstract}  Possibilities of the formation of the pion-sigma meson field vortices in a   rotating empty vessel (in vacuum) and  in the pion-sigma Bose-Einstein  condensates at a dynamically fixed particle number are studied within the linear $\sigma$ model at zero temperature. The charged  $\sigma\pi^{\pm}$ and neutral  $\sigma\pi^0$ complex field ansatze (models 1 and 2) are studied.  First it is analysed,  which of these  configurations is  energetically favorable  in case of the system at rest. Then conditions are found, at which a  chiral field storm can arise in a rotating empty vessel.  In the  model 1  an important role played by  the electric field is demonstrated. Its appearance  may allow for formation of a supervortex (vortex with a large angular momentum)  in case of the  empty vessel  rotating with an overcritical angular velocity. Influence of  magnetic field is also studied.   Field configurations in presence and absence of the meson self-interaction are found in both models. Then it is shown that description  of  the charged (in model 1) and  neutral (in model 2) rotating pion-sigma Bose-Einstein condensates is analogous to that for the Bose-Einstein condensates in cold atomic gases. Various field configurations such as vortex lines, rings and spirals are discussed. Conditions for existence of the  rigid-body  rotation of the vortex lattice   are then  analysed.  Observational effects for vortex fields in rotating vessels,  energetic heavy-ion collisions and in rotating superheavy nuclei and nuclearites are discussed.
\end{abstract}
%\eject
%\date{\today}
 \maketitle
\section{Introduction}
As it is known, the cold $^4$He exists in the superfluid  condensate phase. On the ground of a  ``mother'' uniform condensate the spectrum of excitations has the phonon-roton form with the local minimum of the excitation energy $\omega (k)$ at $\omega (k=k_m)$, $k$ is the excitation momentum. During a long time it was thought that the superfluidity is destroyed completely, if the superfluid moves uniformly with the speed $\vec{W}>u_{\rm L} =\mbox{min} (\omega (k)/k)$, where $u_{\rm L}$ is  the Landau critical velocity,  $\hbar=c=1$.  The criterion
$W<u_{\rm L}$ is usually called the Landau necessary condition of  the superfluidity, cf. \cite{LL9,Tilly-Tilly,PethickSmith,PitString}.
Reference \cite{Pitaev84}  demonstrated that in case of a superfluid $^4$He, undergoing a nonrelativistic rectilinear motion in a capillary with velocity $W>u_{\rm L}$,  there may appear additional ``daughter'' condensate of excitations with
$k=k_m\neq 0$.  Excitations appear owing to a friction  occurring near the walls.  Reference \cite{V93jetp} generalized  consideration to the case of various moving media. %including relativistic systems.
Both uniformly moving and rotating systems were studied.

Reference \cite{Vexp95} considered a possibility of the condensation of the Bose zero-sound-like excitations  with $k_0\neq 0$ in  Fermi liquids moving with $W>u_{\rm L}\simeq v_{\rm F}$, where $v_{\rm F}$ is the Fermi velocity.  In \cite{V93jetp,Vexp95} explicit results were  presented for pions and for various types of zero sounds in moving  nuclear matter. Reference \cite{BaymPethick2012} studied condensation of excitations  with $k_0\neq 0$ (levons) in cold Bose gases.  Then condensation of excitations with $k_0\neq 0$ in moving systems  was considered in \cite{Kolomeitsev:2016isb}.  The results might be applicable for  description of various bosonic sub-systems such as superfluid $^4$He, ultra-cold atomic Bose gases, charged pion and kaon condensates in rotating neutron stars,  various superconducting fermionic systems with the pairing, like proton and color-superconducting components in compact stars, metallic superconductors,  the neutron superfluid component in compact stars, and ultracold atomic Fermi gases, cf. also  \cite{Voskresensky:2022fzk}.
The photon Cherenkov radiation and shock waves in supersonic fluxes (e.g., shock wave appearing  when an airplane overcomes sound velocity) are related phenomena. However in open systems  produced excitations may run away instead of the formation of   the condensate. In case of the single wall moving with the speed $W>u_{\rm L}$ in a medium characterized by appropriate dispersion relation we deal with an open system. In case of two walls
the condensate of excitations can be formed in the region between the walls. In case of the rotating bucket filled by the medium, the condensate of excitations can be formed inside the bucket.

In case of the Bose--Einstein condensation  of nonrelativistic bosons in equilibrium matter one deals with fixed averaged number of bosons determined by the value of their chemical potential, $\mu$, cf. \cite{LL9,Tilly-Tilly,PethickSmith,PitString}.
Besides the daughter condensate of rotonic  excitations, in a vessel filled by the condensed $^4$He rotating with an overcritical frequency $\Omega >\Omega_{c1}\sim \ln R/(mR^2)$ there may appear vortices. Here $m$ is the mass of atoms of He, $R$ is the transverse size of the vessel. Vortices represent  topological  defects keeping the quantum number, the angular momentum.
Formation of vortices in a moving superfluid  is in a sense similar to the mentioned above creation of excitations populating a low-lying branch of the spectrum. Vortices and anti-vortices are produced in pairs. Anti-vortices can be absorbed  at  the walls of the rotating vessel. In this way remaining very massive vortices  select  a part of the angular momentum from the rotating vessel.  At least for buckets of a sufficiently large  size $R$ one has $\Omega_{c1}\ll \Omega_{c\rm L}$, where $\Omega_{c\rm L}\sim u_{\rm L}/R$ is the critical rotation frequency, cf. \cite{V93jetp}, at which there appears condensate of excitations with $k_0\neq 0$. The vortex-rings may also appear at the rectilinear motion of the condensed $^4$He. In this case for systems of a sufficiently large size the critical velocity $W_{c1}$ for formation of vortex-rings proves to be smaller than $u_{\rm L}$. In difference with the daughter condensate of excitations the occurrence of vortices requires presence of a mother condensate. Friction in the mother condensate in presence of vortices and a possibility of formation of  a giant vortex states in rapidly rotating Bose-Einstein condensates were studied in \cite{BaymChandler1983,BaymChandler1986} and \cite{FischerBaym2003}.

Formation of the vortices in the  resting systems is energetically unfavorable \cite{LL9}. Some works considered question about existence   of the vortices in a static matter  within various modifications of the linear $\sigma$ model, cf. \cite{Zhang1997,Berera2016} and references therein.   A role of various topological defects, such as vortex strings in the early Universe, has been   discussed in \cite{Vilenkin2000}.

%%%

At SPS, RHIC and LHC heavy-ion collision energies   at midrapidity a  baryon-poor
medium is formed  \cite{Afanasiev,Alt,Nayak,Abelev,Adamczyk} with pion number exceeding the baryon/antibaryon number more than by the order of magnitude. To describe such a matter authors of the hadron resonance gas model, e.g., cf. Ref. \cite{Stachel}
and references therein, assume that hadrons  are produced at the hadronization temperature
$T_{\rm had}$  after cooling of an expanding quark-gluon fireball.
At the temperature of the chemical freeze-out, $T_{\rm chem}\simeq T_{\rm had}$,  inelastic processes are assumed to be ceased and  up to the thermal freeze out only elastic processes occur. Thus at the stage $T_{\rm chem}>T(t)>T_{\rm th}$ one may speak about a dynamically fixed pion number, i.e. approximately not changing at this time-stage. If the state formed at the chemical freeze-out was   overpopulated, during the cooling there may appear the Bose-Einstein  condensate  of pions   characterized by the dynamically fixed pion number, as it was suggested  in  \cite{Voskresensky1994}. On the time interval, at which the particle number remains almost constant, the Bose-Einstein  condensate  of pions  behaves similar to a superfluid. The problem  was  then studied, e.g.,  in Refs. \cite{Voskresensky:1995tx,KKV1996,
Voskresensky1996,BFR2014,Kolomeitsev2018,Nazarova2019,Blaschke:2020afk,
Kolomeitsev:2019bju}. In \cite{Kolomeitsev2018,Kolomeitsev:2019bju} the  processes keeping the total number of pions approximately fixed were separated from the processes including absorption and production of pions  already in the Lagrangian of the $\lambda (\vec{\Phi}^2)^2$ model.
Similarly to the pion Bose-Einstein  condensation, the Bose-Einstein  condensation of gluons may arise \cite{Blaizot2012,Xu:2014ega,Blaizot2017,Greiner2019,Peshier2019}. A non-equilibrium Bose–Einstein condensation  has been observed and studied  in case of the so-called exiton-polariton bosonic quasiparticles that exist inside semiconductor microcavities  \cite{Exiton2010,Byrnes2014}.
The ALICE Collaboration observed a significant suppression of three and four pion
Bose–Einstein correlations in Pb-Pb collisions at
$\sqrt{s_{NN}} =2.76$ TeV at the LHC \cite{Abelev2014,Adam2016}. This circumstance can be interpreted as there is a considerable degree of coherent pion
emission in relativistic heavy-ion collisions \cite{Akkelin2002,Wong2007}. Analysis \cite{Begun2015} indicated that about $5\%$ of pions could stem from the Bose-Einstein condensate. A discussion of a possibility of the  Bose-Einstein condensation  in heavy-ion collisions at LHC energies can be found in the reviews \cite{Shuryak2017,Voskresensky:2022fzk}.

Estimates show presence  of angular
momenta $L\sim \sqrt{s}Ab/2\lsim 10^6\hbar$ in peripheral heavy-ion collisions of Au $+$ Au  at $\sqrt{s} = 200$ GeV, for impact parameter $b = 10$ fm, where $A$ is the nucleon number of the ion, cf.  \cite{Xu-Guang Huang}. The global
polarization of $\Lambda (1116)$  hyperon observed  by the STAR Collaboration in non-central gold-gold collisions  \cite{Adamczyk} indicated on existence of a vorticity with  rotation frequency  $\Omega\simeq
(9\pm 1) \cdot 10^{21}$ Hz $\simeq 0.05m_\pi$.

Besides a rotation, also strong magnetic fields are expected to occur at heavy-ion collisions and in compact stars. Already first estimates \cite{Voskresensky:1980nk} predicted values of the magnetic field up to  $\sim (10^{17}-10^{18})$G  for peripheral heavy-ion collisions at the energy $\sim$ GeV per nucleon. Also fields with $H\lsim (10^{15}-10^{16})$G exist at surfaces of magnetars and  with  $H\lsim 10^{18}$G may be  may exist in  neutron star interiors. Inhomogeneous pion condensation, which may appear in a dense nucleon matter, may survive even in such strong magnetic fields \cite{Voskresensky:1980nk}. Question about condensation of the noninterating charged pions in vacuum at a  simultaneous action of the rotation and a sufficiently strong magnetic field was  studied in  \cite{Zahed}.
Reference \cite{Guo} included the pion self-interaction within the $\lambda|\Phi|^4$ model and suggested  appearance of a giant pion vortex state (supervortex). Effects of electric field  were disregarded.

The  chiral SU(2)$\times$ SU(2)  symmetry, which is isomorphic to the symmetry under rotations in Euclidean 4-space-time,  plays an  important role in the nuclear phenomena involving  hadron interactions and in the strong interaction theory based on quark-gluon degrees of freedom. This symmetry holds for massless particles and it is broken by the terms in the Lagrangian  associated with particle masses.
The quarks $u$ and $d$ have very small bare masses compared with the nucleon mass $m_N\simeq 938$ MeV, namely $m_u, m_d$ are estimated as $\simeq (2-5)$ MeV. Experimental value of the pion mass $m_\pi\simeq 140$ MeV is also small compared to $m_N$. If one puts  $m_u =0$, $m_d =0$ and $m_\pi =0$, the model becomes  SU(2)$\times$ SU(2)  symmetric.  In the vacuum the symmetry is spontaneously broken and the $\sigma$ field acquires the non-zero expectation value. Within the hadronic sigma model the nucleons  are  introduced  as initially massless particles and their large experimental value of the mass   appears then as the result of the interaction of the nucleon and the vacuum $\sigma$ meson field. Small pion mass term can be then added,  as a  term explicitly breaking the SU(2)$\times$ SU(2)  symmetry.

In the given paper within the linear sigma model we employ two models, one (model 1) describing complex charged pion field and another one (model 2) describing neutral complex $\sigma\pi^0$ field. The latter model  allows to consider fluidity in the system of electrically neutral $\sigma\pi^0$ field. First we study conditions, at which either solutions of  the  model 1 or of the  model 2 are energetically favorable in the system at rest. Then we consider rotating systems. First,  the question will be studied, whether  the chiral-field vortices may appear  in the rotating empty vessel (in the vacuum) in absence and in presence of the static   electric potential. Then  we study Bose-Einstein condensates of the charged pions in model 1 and of $\sigma\pi^0$ in model 2  in the rotating vessel    and in a piece of the hadron  matter. In the case of the rotating gases with a dynamically fixed particle number  the vortices may form the lattice.
We will consider a  possibility to seek for various observational consequences of the formation of  chiral-vortex structures. Recently a GigaHerz-frequency rotation of the dynamical exiton-polariton Bose-Einstein condensates has been studied in \cite{Redondo2023}.

The paper is organized as follows. Section \ref{Linear-sect} introduces the Lagrangian of the linear $\sigma$ model. Two specific  models are considered, one corresponding to a possibility of the  formation of the charged pion field and another one to a possibility of the  formation of the  neutral $\sigma\pi^0$ complex field in the system  at rest. Section \ref{chir} studies chiral fields, which can be formed within mentioned two models in the rotating systems. First we consider the case of not self-interacting bosons and then switch on the self-interaction. We study   possibilities of the formation of the chiral-field vortex condensates in an empty rotating vessel (vacuum)  in absence and in presence of the static electric potential, an artificial external vector field and the magnetic field. Then we consider appearance of a vortex field  in   a rotating piece of nuclear matter and in the  rotating  gases with a dynamically fixed particle number. Section \ref{concluding-sect}
contains conclusions. In this paper we will use units $\hbar=c=1$.

\section{Linear sigma model}\label{Linear-sect}
\subsection{Zero pion mass}
The linear sigma model as realization of the chiral symmetry  was introduced by J. Schwinger in 1957 and Gell-Mann and Levy in 1960, cf. \cite{Lee72}. In this  model the isospin triplet of pseudoscalar pions $\vec{\pi}= (\pi_1, \pi_2,\pi_3)$ is unified with the scalar meson $\sigma$ in the Euclidean quadruplet $\phi_\alpha = (\sigma, \pi_1, \pi_2,\pi_3)$, $\alpha =1,2,3,4$, $\phi^2_\alpha =\sigma^2 + \vec{\pi}^{\,2}$.
In absence of nucleons the Lagrangian density is as follows
\be
{\cal{L}}_\phi =\frac{\partial_\mu \phi_\alpha \partial^\mu \phi_\alpha}{2}-\frac{\lambda (\phi^2_\alpha -v^2)^2}{4}\,.\label{Lphi}
 \ee
This Lagrangian is symmetric under rotations in 3-isospin space of fields $\pi_i$  and 4-space of fields $\phi_\alpha$. Comparison of different terms in (\ref{Lphi}) shows that the fields $\sigma$, $\pi_i$ and constant  value  $v$ have  dimensionality $1/l$, where $l$ is a length scale, and constant $\lambda$ is dimensionless.
We assume that $\lambda$ and $v^2$ are positive constants. Positiveness of $\lambda$ is needed for  stability of the vacuum. Positiveness of $v^2$ provides spontaneous chiral symmetry breaking.

It is convenient to introduce  complex chiral fields in the form
\be\Phi_{\sigma} =(\sigma +i\pi_3)\,,\quad  \Phi_\pi =(\pi_1 +i\pi_2)=\sqrt{2}\varphi\,.\label{Phipi}
\ee
From (\ref{Lphi})  we recover equations of motion
\be(-\partial_t^2+\Delta) \Phi_{\sigma,\pi} +[\lambda v^2  -\lambda (|\Phi_\sigma|^2+|\Phi_\pi|^2)]\Phi_{\sigma,\pi} =0.\label{piSigma}
\ee
In the uniform space this equation has constant solution  $|\Phi_\sigma|^2+|\Phi_\pi|^2 =v^2$.

In spite of the symmetry, the vacuum  is usually described by  $\sigma =\pm v$ (to be specific we further fix  $\sigma =+ v>0$) and $\pi_i=0$ because the choice, e.g., $\pi_1 =v$ (or $\pi_2 =v$, or $\pi_3 =v$) would spoil isotopic symmetry of strong interactions. Setting $\Phi_\sigma =v +\Phi'_\sigma$, $\Phi_\pi =0$ in Eq. (\ref{piSigma})  we arrive at equation \be(-\partial_t^2+\Delta) \Phi'_\sigma -\lambda v^2 (\Phi'_\sigma+\Phi^{*\prime}_\sigma) +O(\Phi^{\prime 2}_\sigma)= 0\,.\label{pionqpSigmamass}\ee
From (\ref{pionqpSigmamass}) and the corresponding equation for $\Phi^{*\prime}_\sigma$ it follows that
the neutral $\sigma' =(\Phi'_\sigma+\Phi^{*\prime}_\sigma)/2$ excitation acquires the mass term
\be m_{\sigma}=\sqrt{2\lambda v^2}>0\,,\label{Phimass}\ee
whereas the  pion excitations remain massless. Employing $m_\sigma\simeq 600$ MeV we have  $\lambda \simeq 20$. With the choice $\Phi_\sigma =0$, $\Phi_\pi \neq 0$  one would have  $m_{\pi_1}= \sqrt{2\lambda v^2}>0$ (or $m_{\pi_2}= \sqrt{2\lambda v^2}>0$) and massless other fields, that is not realistic.

 The conserved 4-current densities associated with the $\Phi_\sigma$ and $\Phi_\pi$ fields are as follows
\begin{eqnarray} &j^\nu_\Phi =(\rho_\Phi, \vec{j}_\Phi)=-(\Phi_\sigma^*\partial^\nu\Phi_\sigma -\Phi^*_\sigma\partial^\nu\Phi_\sigma)/(2i)\,,\nonumber\\
&{j}^\nu_\pi =-(\Phi_\pi^*\partial^\nu\Phi_\pi -\Phi^*_\pi\partial^\nu\Phi_\pi)/(2i)\,,\label{j}\end{eqnarray}
and with the stationary field ansatze $\Phi_\sigma \propto e^{-i\mu_\Phi t}$, $\Phi_\pi \propto e^{-i\mu_\pi t}$, we obtain
\be\rho_\Phi =\mu_\Phi |\Phi_\sigma|^2\,,\quad \rho_\pi =\mu_\pi |\Phi_\pi|^2\,.\label{muPhi}\ee
Here, the chemical potential $\mu_\Phi$   relates to the conserved axial current
\be
A^\mu_j =%\frac{1}{2}\bar{N}\gamma^5 \gamma^\mu \tau_i N +
\pi_j \partial^\mu \sigma - \sigma \partial^\mu \pi_j \,,\ee
$j=1,2,3$,  and
 the quantity $\mu_\pi$ is the electro-chemical potential associated with conservation of the electromagnetic current.
The Lagrangian- and energy densities render
\begin{eqnarray} &{\cal{L}}=\frac{\mu^2_\Phi |\Phi_\sigma|^2}{2}-\frac{|\nabla \Phi_\sigma|^2}{2}-\frac{\lambda (|\Phi_\sigma|^2 +|\Phi_\pi|^2 -v^2)^2}{4}\nonumber\\
&+\frac{\mu^2_\pi|\Phi_\pi|^2}{2}-\frac{|\nabla \Phi_\pi|^2}{2}\,\label{LPhipi}\end{eqnarray}
and
\begin{eqnarray}
&E=\mu_\Phi\rho_\Phi +\mu_\pi\rho_\pi -{\cal{L}}
%%%\nonumber\\&=\frac{\mu^2_\Phi|\Phi_\sigma|^2}{2}+\frac{|\nabla \Phi_\sigma|^2}{2}+\frac{\lambda (|\Phi_\sigma|^2 +|\Phi_\pi|^2 -v^2)^2}{4}+\frac{\mu^2_\pi|\Phi_\pi|^2}{2}+\frac{|\nabla \Phi_\pi|^2}{2}
\,.\label{EnerPhipi}\end{eqnarray}
The minimum of the energy  corresponds to  $\nabla \Phi_\sigma =\nabla \Phi_\pi =0$,  and $\mu_\Phi =\mu_\pi =0$, i.e. to the absence of particles in the vacuum. Then  $E=0$ and $|\Phi_\sigma|^2 +|\Phi_\pi|^2 =v^2$. At a fixed particle number the particle density, Eq. (\ref{muPhi}), is found from (\ref{LPhipi}) as $\rho_{\pi,\Phi}=\partial {{\cal L}}/\partial \mu_{\pi,\Phi}$. The same is correct in case of a dynamically fixed particle number provided one considers the system during the time interval shorter than the decay time.
In the description of ultrarelativistic heavy-ion collisions at a late collision stage (after the so called chemical freeze out till the thermal freeze out) inelastic pion-pion processes are assumed to be suppressed compared to elastic scatterings and the total pion number can be considered as approximately fixed, whereas processes $\pi^{+}+\pi^-\leftrightarrow 2\pi^0$ are allowed  \cite{Kolomeitsev:2019bju}.
Then the quantities  $\mu_\Phi$ and $\mu_\pi$ are determined from the condition of  the fixed particle number.

To study further effects of rotation we need to deal with complex fields.
 For this aim to avoid extra complications we will  consider two models. \\{\em Model 1}: $\sigma =v, \pi_3 = 0$, and $\pi_1\neq 0$ or/and  $\pi_2\neq 0$. This model  permits to study fluidity of the $\pi^{\pm}$ charged fields. Note that with $\Phi_\sigma =0$ the Lagrangian (\ref{LPhipi}) coincides with the Lagrangian of the ordinary $\lambda |\Phi_\pi|^4/4$ model. \\
{\em Model 2}: The mean field $\sigma\neq 0$ or/and $\pi_3\neq 0$, whereas $\pi_1=\pi_2=0$, $\sigma^2+\pi_3^2=v^2$. This model allows to consider fluidity in the neutral $\sigma, \pi^{0}$ system. In the limiting case $\sigma =v$, $\Phi_\pi =0$ the Lagrangian density (\ref{LPhipi}) coincides with the Lagrangian density of the  $\lambda \pi_3^4/4$ model. Previously the ansatz of the model 2, which allows to describe a homogeneous condensate of the $\sigma\pi^0$  complex electrically neutral field, has  been considered within the NJL model.

A remark is in order. The $\pi^-$ and $\pi^+$ behave differently in respect to the electromagnetic interactions.  For a positively charged nuclear droplet of a fixed charge density $\rho^{\rm ch}$ the Coulomb energy grows with the radius of the system ($Z=4\pi R^3\rho^{\rm ch}/3$) as ${\cal{E}}\propto Z^{5/3}$, whereas the strong interaction part of the energy is $
\propto Z$. Thereby for nuclear systems of a rather small size $R\sim Z^{1/3}$ the Coulomb effects  can be neglected, whereas they  become efficient at the increase of $Z$.
For a sufficiently large $Z$, the $\pi^-$ level reaches zero and the  nuclear droplet with approximately the same number of protons and neutrons  via reactions $n\to p+\pi^-$  produces the $\pi^-$ condensate  and the  interior becomes to be electro-neutral, cf. \cite{Migdal:1977rn,Voskresensky:1978uf}.

 \subsection{Non-zero pion mass}\label{breaking-sect}
As we have mentioned, in order to describe that   pions have a  non-zero physical mass $m_\pi \simeq 140$ MeV $\ll m_N\simeq 934$\,MeV, $m_N$ is the nucleon mass,  one introduces  the term in the Lagrangian density (\ref{Lphi}), which explicitly breaks the chiral symmetry. One  may use the choices:
\be
{\cal{L}}_{\rm s.b.}^{(1)}=\epsilon \sigma \,\quad{\rm or}\quad  {\cal{L}}_{\rm s.b.}^{(2)}=-\frac{m^{*2}_\pi \vec{\pi}^2}{2}\,.\label{Lsb1}
 \ee
 The value $\epsilon$ has  sense of the constant density of the scalar charge.
As it is explicitly seen, the term  ${\cal{L}}_{\rm s.b.}^{(2)}$ describes the massive pion. Here we use the value $m^*_\pi$ instead of $m_\pi$, that may be particularly useful
for comparison with the lattice gauge theory. Also some authors, cf. \cite{Bowman2009}, studying general properties of the phase diagram in the $\sigma$ model allowed for a variation of the value $m^*_\pi$. Then an interesting phase structure occurs, which results in zero, one, or two critical
points depending on the value of $m^*_\pi$. Moreover, the nucleon-pion interaction in the baryon matter results in appearance of the pion energy-momentum dependent effective pion mass. Such an attractive interaction allows for a p-wave (and s-wave in some models) pion condensation in a sufficiently dense baryon matter, cf. \cite{Migdal1978,MSTV90,Voskresensky:2022gts}. In case of hypothetical Bose stars \cite{Schunck}
the gravitational potential in the nonrelativistic limit is added to the value $m_\pi^2$.

To be specific  we  further  employ  the   symmetry breaking term in the form ${\cal{L}}_{\rm s.b.}^{(2)}$, which  explicitly demonstrates  appearance of the pion mass.  With the term ${\cal{L}}_{\rm s.b.}^{(1)}$  at hand it is still necessary to show that the pion acquired  mass. This case is discussed in Appendix 1.

\subsection{Charged and neutral pion-sigma condensates}

Taking $\Phi_\pi=\Phi_{0\pi} e^{-i\mu_\pi t}$, $\sigma =v$ in model 1 and $\Phi_\sigma=\Phi_{0\sigma} e^{-i\mu_\Phi t}$ in model 2 we rewrite the Lagrangian density (\ref{LPhipi}) as
\begin{eqnarray} &{\cal{L}}_{\pi,\Phi}^{V}=\frac{(\mu_\Phi-g\omega_0)^2|\Phi_{0\sigma}|^2}{2}-\frac{|\nabla \Phi_{0\sigma}|^2}{2}-\frac{\lambda (|\Phi_{0\sigma}|^2 +|\Phi_{0\pi}|^2 -v^2)^2}{4}\nonumber\\
&+\frac{(\mu_\pi -V-g\omega_0)^2|\Phi_{0\pi}|^2}{2}-\frac{|\nabla \Phi_{0\pi}|^2}{2}-\frac{m^{*2}_\pi \vec{\pi}^2}{2}\nonumber\\
&+\frac{(\nabla V)^2}{8\pi e^2}+n_p V\,.\label{phiV}\end{eqnarray}
Here we added a contribution of the static electric field $V=eA_0$ in case of the model 1 describing charged particles, $n_p$ is the density of an ``external'' charge, $e$ is the electron charge. If $n_p\neq 0$, the field $V$ appears even in absence of the field $\Phi_\pi$. For instance, the charged density can be associated with a proton distribution,  if we deal with a piece of the nuclear matter,  or with a  surface charge  placed on plates of the capacitor, if the system is placed in a capacitor. In case of a cylindrical co-axial capacitor, in absence of the charged boson field the electric potential is constant in the region inside the inner cylinder.  In case of the model 2 we should put $V=0$. For generality we also added an interaction of our scalar complex field with the zero-component of an external neutral vector field, e.g., the $\omega_\mu$ vector-meson field, $g$ is the corresponding coupling constant, and  we will  consider the case  $g\omega_0\simeq const <0$. Unfortunately, the value of the coupling $g$ is not experimentally constrained and ordinary one puts it zero.
Inclusion of the $\sigma$-pion-nucleon interaction will be performed in  subsection \ref{sect-nucleon-pion}.

\subsubsection{ Complex charged pion field} In the model 1 equation of motion for the  charged pion field is
   \begin{eqnarray}
\Delta \Phi_{0\pi}   +[(\mu_\pi-V-g\omega_0)^2-m^{*2}_\pi -\lambda |\Phi_{0\pi}|^2]\Phi_{0\pi} =0.
\label{piSigma1}
\end{eqnarray}

By variation of the action over $V$ we recover the Poisson equation
\be \Delta V=4\pi e^2 (n_p +(V+g\omega_0-\mu_\pi)|\Phi_{0\pi}|^2)\,.\label{Poisson}
\ee
In a broad electric potential well, $Rm_\pi^*\gg 1$, the ground-state $\pi^-$ energy level dives into the lower continuum, $\mu_\pi \leq -m_\pi^*$,  for $V<V_c=-2m_\pi^*$  at $g\omega_0=0$. In case of the plain capacitor, $n_p$ is  the charge density distributed on the plates. The critical value $V_c =eEl=-2m_\pi$ can be reached at fixed value of the electric strength   $E\sim 10^5$v$/$cm  by pushing  the plates of the capacitor apart on the distance  $l\sim 10$m. However  $\pi^{\pm}$ pairs are produced in the tunneling process and the probability of their creation at such low electric fields is negligibly small $W\sim e^{-\pi m_\pi^2/(|e E|)}$ and thereby the process needs too long time to be observed, cf. \cite{V88,Voskresensky2021QED}.

In case of approximately uniform system for $V\simeq -V_0=const$ from Eq. (\ref{piSigma1})  we recover
\begin{eqnarray} &|\Phi_{0\pi}|^2=[(\bar{\mu}_\pi^2 -m_\pi^{*\,2})/\lambda]\theta (\bar{\mu}^2_\pi -m_\pi^{*\,2})\,,\nonumber\\
&\bar{\mu}_\pi=\mu_\pi +V_0-g\omega_0\,,\label{pionphicond}
\end{eqnarray}
where we introduced the value of the shifted chemical potential, $\theta(x)$ is the step-function.

The case $V_0\simeq 0$ is relevant for consideration of nuclear systems of a not too large size, e.g.  such as light nuclei. Indeed in  this case  the Coulomb effects prove to be much weaker than the strong-interaction ones and $l_V\sim R$, where $l_V$ is the typical length of the change of the electric field. On the other hand, for the system of a large size $R\gg l_\pi, l_V$, where $l_\pi$ is the   typical length of the change of the charged pion condensate field, for the case $n_p=0$ the global electroneutrality condition  $\int_0^R \rho (r) r dr =0$ should be satisfied yielding the averaged value of the electric potential $\bar{V}=\mu_\pi -g\omega_0$.

For $n_p (r)=const>0$ at $r<R$ and 0 for $r>R$, we  can put  $V(r)=-V_0=const$ everywhere at $r<R-l_V$, i.e.,  except a narrow region near the edge, cf. \cite{Migdal:1977rn,Voskresensky:1978uf,Voskresensky:1977mz,
%Voskresensky1977,
V90}.

As it follows from (\ref{EnerPhipi}), (\ref{phiV}) for $\lambda\neq 0$ at $V=-V_0\simeq const$ the energy density is given by
\begin{eqnarray}
&E_{\pi}^{V}\simeq {\mu}_\pi \rho_\pi -\frac{(\bar{\mu}_\pi^2-m_\pi^{*\,2})^2}{4\lambda}+n_p V_0\,,\nonumber\\
 &\rho_\pi =\bar{\mu}_\pi (\bar{\mu}_\pi^2-m_\pi^{*\,2})/\lambda \,,\label{Edynpi}
\end{eqnarray}
for $\bar{\mu}^2_\pi >m_\pi^{*\,2}$ and  for $\rho_\pi \ll m_\pi^{*\,3}/\lambda$ we obtain
\begin{eqnarray}
&\bar{\mu}_\pi \simeq m_\pi^{*}+\rho_\pi \lambda/(2m_\pi^{*\,2})\,,\nonumber\\ &E_{\pi}^{V}\simeq m_\pi^{*}\rho_\pi + \frac{\lambda \rho_\pi^2}{4m_\pi^{*\,2}}+(n_p -\rho_\pi) V_0+g\omega_0\rho_\pi\,.
\label{Edynpiappr}\end{eqnarray}

 If we deal with a piece of the  isospin-symmetric nuclear matter, the chemical potential $\mu_\pi$ may reach zero, as it follows from the condition of the equilibrium  respectively the nuclear reactions $n\leftrightarrow p+\pi^-$. For $\mu_\pi =0$, $n_p\simeq \rho_\pi$,  we have
\be
E_{\pi}^{V}(\lambda\rho_\pi\ll m_\pi^{*\,3}, \mu_\pi =0)\simeq  (m_\pi^{*}+g\omega_0) n_p\,.\label{Epilambda0}
\ee
The total energy density  renders \cite{Migdal:1977rn}
 \be {{ E}}=E_{\pi}^{V} +E_{ A}\,,\ee
where $E_{A}$ is the strong interaction contribution and $E_{\pi}^{V}$ is given by Eq. (\ref{Edynpiappr}), (\ref{Epilambda0}). Considering limiting case when  $\lambda$  is very small and $g\omega_0=0$, from Eq. (\ref{piSigma1}) we find that $V_0=m_{\pi}^{*}$ (provided  $V_0 \sim Ze^2/R\gsim m_{\pi}^{*}$).
For the  nucleus of a large atomic number $A$  and for $\rho_\pi =n_p=\rho_0/2$, where $\rho_0$ is the nucleus saturation density, the strong interaction contribution is ${\cal{E}}_{A}\simeq -32Z$ MeV, $Z\simeq A/2$ is the proton charge of the nucleus. For $\lambda \to 0$ we would have
\be {\cal{ E}}\simeq (m^{*}_\pi -32 \mbox{MeV})Z \label{freesupercharged}
\ee
and we would deal with  stable $\pi^-$ condensate nuclei and nuclei-stars  of arbitrary size (till their atomic number $A$ is  $\ll 10^{57}$ and effects of the gravity can be neglected),  if  $m^{*}_\pi$ were smaller than 32 MeV, cf.  \cite{Voskresensky:1978uf,Voskresensky:1977mz,MSTV90}.

For $\lambda\neq 0$, $\rho_\pi\gg m_\pi^{*\,3}/\lambda$ we get
$$\bar{\mu}_\pi \simeq \lambda^{1/3}\rho_\pi^{1/3}(1+m_\pi^{*\,2}/(3\lambda^{2/3}\rho_\pi^{2/3}))$$
 and
\begin{eqnarray} &E_{\pi}^{V} (\rho_\pi\gg m_\pi^{*\,3}/\lambda)\simeq 3\lambda^{1/3} \rho_\pi^{4/3}/4
%-\rho_\pi^{2/3}m_\pi^{*\,2}/\lambda^{1/3}
+\frac{m_\pi^{*\,2}\rho^{2/3}_\pi}{6\lambda^{1/3}}\nonumber\\
&+(n_p -\rho_\pi)V_0+g\omega_0 n_p\,,\label{EdynpiLarge}
\end{eqnarray}
where the first term is dominant. In case of a piece of the  isospin-symmetric nuclear matter, which we have discussed,   taking $\lambda \simeq 20$, $m^{*}_\pi=m_\pi$,  we may use Eq.
%(\ref{Edynpiappr}),
(\ref{EdynpiLarge}).
%{EdynpiLarge}).
Then we have
\be {\cal{ E}}\simeq (3\lambda^{1/3}\rho_\pi^{4/3}/4  +m_\pi -32 \mbox{MeV})Z\simeq  m_\pi Z>0\,.\label{selfsupercharged}\ee
Thereby a decrease of the energy due to the charged pion condensation at ignorance of the pion-nucleon interaction is not sufficient for  existence of stable approximately isospin-symmetric  nuclei of a large size with the nucleon density $\rho\simeq\rho_0$ in their interiors, at the positive  charge  compensated by the negative charge of the pion condensate field, cf.  \cite{Voskresensky:1978uf,Voskresensky:1977mz,MSTV90}.  Influence of the pion condensation with the momentum $k\neq 0$, which can be formed at a density $\rho>\rho_c^\pi>\rho_0$ owing to a strong p-wave pion-nucleon interaction, on a possibility of existence of supercharged superheavy  nuclei was studied in  \cite{Voskresensky:1977mz,
%Voskresensky1977,
MSTV90,V88,Voskresensky:2020shb,Gani:2018mey}.

\subsubsection{ Neutral complex pion-sigma field} In the model 2    the time averaging of the term $-{m_\pi^{*\,2}\pi_3^2/2}$ can be presented as
$-{m_\pi^{*\,2}\Phi_{0\sigma}^2/4}$, since $\pi_3=\Phi_{0\sigma}\sin (\mu_\Phi t)$, as it follows from Eq. (\ref{Phipi}).
Equation of motion for $\Phi_\sigma$ is then given by
    \begin{eqnarray}
    (\bar{\mu}_\Phi^2+\Delta) \Phi_{0\sigma} +[\lambda (v^2 -|\Phi_{0\sigma}|^2 ) -m_\pi^{*\,2}/2]\Phi_{0\sigma}   =0\,,\label{piSigma111}
  \end{eqnarray}
where
$\bar{\mu}_\Phi =\mu_\Phi -g\omega_0$.

In  case of approximately uniform system from (\ref{piSigma111}) we find the mean-field  solution
\begin{eqnarray}&\Phi_{0\sigma}^2= [v^2 +\
{(\bar{\mu}_\Phi^2 -m_\pi^{*\,2}/2)}/{\lambda}]\nonumber\\
&\times\theta(v^2 +{(\bar{\mu}_\Phi^2 -m_\pi^{*\,2}/2)}/{\lambda})\,,\quad \Phi_{0\pi} =0\,, \label{phisol}\end{eqnarray}
where $\theta (x)$ is the step-function.

 The energy density is as follows
\begin{eqnarray}
&E_\Phi =\bar{\mu}_\Phi \rho_\Phi +g\omega_0 \rho_\Phi -\frac{(\bar{\mu}_\Phi^2 - m_\pi^{*\,2}/2)v^2}{2}\nonumber\\&- \frac{(\bar{\mu}_\Phi^2 - m_\pi^{*\,2}/2)^2}{4\lambda}\,.\label{genEPhi}
\end{eqnarray}
For $\lambda\rho_\Phi \ll m_\pi^{*3}$, $\lambda \ll 1$ we have $\bar{\mu}_{\Phi}\simeq
m_\pi^{*} (1-\lambda v^2/m_\pi^{*2}+...)/\sqrt{2}$  and $E_\Phi (\lambda\rho_\Phi \ll m_\pi^{*3})=\lambda v^4/4+\rho_\Phi m_\pi^{*} /\sqrt{2}+...$.

We see that for $n_p=0$, $g\omega_0=0$ Eq. (\ref{Edynpiappr}) yields a smaller energy than (\ref{genEPhi}).

In a physically meaningful case  $\lambda \gg 1$,  for $ \rho_\Phi \ll v^3\sqrt{\lambda}$, as it follows from relations (\ref{muPhi}) and (\ref{phisol}), we obtain
\begin{eqnarray} &E_\Phi
\simeq \frac{\rho_\Phi^2}{2v^2}+\frac{m_\pi^{*\,2}v^2}{4}+g\omega_0 \rho_\Phi+O({1}/{\lambda})\,,\nonumber\\
&\bar{\mu}_\Phi \simeq \rho_\Phi/v^2(1+O(1/\lambda))\,.
\label{EPhidyn}\end{eqnarray}
For $\rho_\Phi \gg m_\pi^{*}v^2$ the first  term is dominant. For $\rho_\Phi \ll m_\pi^{*}v^2$ the second  term is dominant.

For  $\rho_\Phi \gg v^3\sqrt{\lambda}$ one has
\begin{eqnarray} &E_\Phi\simeq \frac{3\lambda^{1/3} \rho_\Phi^{4/3}}{4} -\frac{(\lambda\rho_\Phi)^{2/3}v^2}{2}\left[1+O\left(\frac{1}{\lambda}\right)\right]
%-\frac{m_\pi^{*\,4}}{16\lambda}
\,,\,\, \label{largelimphi}\\
&\bar{\mu}_\Phi \simeq \lambda^{1/3}\rho_\Phi^{1/3}\left(1-\frac{\lambda^{1/3}v^2}{3\rho_\Phi^{2/3}}\right)\,.
\nonumber\end{eqnarray}

\subsubsection{ Energetical favorability of various configurations}
As we see from the above derived expressions, for $n_p=0$, $g\omega_0=0$, i.e. in absence of external fields, the energetically favorable solutions correspond to $\rho_{\pi,\Phi}=0$ and solution in the model 2 is energetically unfavorable, whereas the solution in the model 1 describes the vacuum state $\sigma =\pm v$, $\pi_i=0$. In case of a pion gas with fixed particle number, $\rho_{\pi,\Phi}\neq 0$,  the system  energy given  by Eqs. (\ref{Edynpi}) and (\ref{genEPhi}) is positive for both models 1 and 2. The case $\lambda =0$ is specific. Here we have $\bar{\mu}_{\pi}={m}^*_\pi$ in the model 1 and $\bar{\mu}_{\Phi}={m}^*_\Phi/\sqrt{2}$ in the model 2, $\Phi_{\pi,\Phi}$ are expressed through  $\rho_{\pi,\Phi}$ according to  Eqs. (\ref{muPhi}) after performing the replacements ${\mu}_{\pi}\to \bar{\mu}_{\pi}$ and ${\mu}_{\Phi}\to \bar{\mu}_{\Phi}$.

For $n_p=0$, $g\omega_0=0$ in the case of {\em the dynamically fixed pion number} for a system heaving a rather small size, when we may put $V_0\to 0$,
comparing (\ref{EPhidyn}) and (\ref{Edynpiappr}) at fixed density $\rho_\pi=\rho_{\pi^+}+\rho_{\pi^-}=\rho$   in the model 1 and
$\rho_{\pi^{0}}=\rho_\Phi/2 =\rho$ in the model 2  we see that  for  $\rho_{\pi}\ll m_\pi^{*\,3}/\lambda <m_\pi^{*}v^2<v^3\sqrt{\lambda}$ the value (\ref{Edynpiappr}) is smaller then (\ref{EPhidyn}), i.e. solution of the model 1 is energetically favorable compared to that for the model 2.
The same statement holds for $\rho_{\pi^{+}}=\rho_{\pi^{-}}=\rho_{\pi}/2=\rho_{\pi^{0}}$.
 In case $\rho >v^3\sqrt{\lambda}$, for  $\rho_{\pi^+}+\rho_{\pi^-}=\rho_{\pi^{0}}=\rho$ again solution of the model 1 is energetically favorable compared to that for the model 2. However for   $\rho_{\pi^{+}}=\rho_{\pi^{-}}=\rho_{\pi}/2=\rho_{\pi^{0}}$  Eq. (\ref{largelimphi}) leads to a smaller energy than (\ref{EdynpiLarge}), i.e. solution of the model 2 for $\rho >v^3\sqrt{\lambda}$ is energetically favorable compared to that for the model 1. Please notice that  Refs. \cite{Kolomeitsev2018,Kolomeitsev:2019bju} demonstrated that initially isospin-asymmetric pion gas with a dynamically fixed total particle number at a   temperature $T>T_{\rm BEC}$  by reactions $2\pi^0\leftrightarrow \pi^+\pi^-$ will form the state with
 $\rho_{\pi^{-}}=\rho_{\pi}/2=\rho_{\pi^{0}}$, where $T_{\rm BEC}$ is the critical temperature of the Bose-Einstein condensation. The possibility of the $\sigma\pi^0$ condensate fields at $T=0$ was not considered there.

In an artificial case,  $m_\pi^{*}\to 0$ (chiral limit),  for $\rho_{\pi^{0}}\ll v^3\sqrt{\lambda}$
we find that
Eq. (\ref{EPhidyn}) yields a smaller energy than (\ref{EdynpiLarge}), i.e. solution in the  model 2 is energetically favorable compared to that in the model 1. For $\rho\gg v^3\sqrt{\lambda}$, comparing (\ref{largelimphi}) and (\ref{EdynpiLarge}) we see that the solution in the model
1 is energetically favorable compared to that in  the model 2 for $\rho_\pi=\rho=\rho_{\pi^{0}} =\rho_\Phi/2$
and the solution in the model
2 becomes to be energetically favorable compared to that for the model 1 for $\rho_{\pi^{+}}=\rho_{\pi^{-}}=\rho_{\pi}/2=\rho_{\pi^{0}}=\rho_\Phi/2$.

In case of {the zero Bose particle number} and $n_p=0$, $g\omega_0=0$ we should put $\mu_\Phi=\mu_\pi =0$, $V_0=0$. In the {model 1} from (\ref{piSigma1}) we have
\be
E_\pi (\mu_\pi=0) =0\,, \quad \sigma =\pm v,\quad \pi_i =0.\label{mu0solpi}
\ee
In the { model 2} from (\ref{genEPhi}) we get
\be
E_\Phi (\mu_\Phi=0)=\frac{m_\pi^{*\,2}v^2}{4}-\frac{m_\pi^{*\,4}}{16\lambda}>0\,,
\label{mu0sol}
\ee
for $|\Phi_\sigma|^2=v^2$, $\Phi_\pi =0$, i.e. solution   (\ref{mu0solpi}) of the model 1 is energetically  favorable in comparison with solution (\ref{mu0sol}) of model 2. Thus the vacuum in the rest frame is stable respectively formation of the charged and neutral pion fields.

Finally let us notice that we could consider one more case: $\sigma =\pi_3=0$, $\Phi_\pi =\pi_1+i\pi_2\neq 0$. In this case  we would have
\be |\Phi_{0\pi}|^2=[v^2+(\mu^2_\pi-m_\pi^{*\,2})/\lambda]\theta(v^2+(\mu^2_\pi-m_\pi^{*\,2})/
\lambda) ,\nonumber\ee
and
$\sigma$ is massless. For a low pion density this possibility is  not realistic.
However in presence of  a large  nucleon density
a developed  p-wave pion condensate may appear  resulting in the chiral transition from the $\sigma \simeq v$ vacuum  to the $|\pi|^2 \simeq v^2$ one, cf. \cite{Baym1975,VM1982}.

\subsection{Nucleon-meson interaction}\label{sect-nucleon-pion}
With inclusion of nucleons the Lagrangian density of the model is  as follows
\begin{eqnarray}
&{\cal{L}}_{\rm tot} =\frac{\partial_\mu \phi_\alpha \partial^\mu \phi_\alpha}{2}-\frac{\lambda (\phi^2_\alpha -v^2)^2}{4}\nonumber\\
&+\bar{N}i\gamma^\mu \partial_\mu N -g\bar{N}(\sigma +i\vec{\tau} \vec{\pi}\gamma^5)N +{\cal{L}}_{\rm s.b.}^{(i)}\,,\label{LphiN}
\end{eqnarray}
$i=1,2$, $\gamma^\mu$ are the Dirac matrices, cf. \cite{Lee72}. Interaction with a static electric field can be introduced similar to that done in Eq. (\ref{phiV}).

In the case $v^2<0$,  ${\cal{L}}_{\rm s.b.}^{(i)}=0$, the model describes massless nucleons, and the pions and the sigma mesons of the same mass $m^2_\sigma = m^{*2}_\pi
=-\lambda v^2>0$. Such a theory does not satisfy the data, which show that nucleons are very massive particles and  pions are  lightest among hadrons. Thus,  one should take $v^2>0$ in the second term of Eq. (\ref{LphiN}), as we have done it above.
Without loosing generality  we may chose $\sigma =v$ and introduce fields describing excitations over the vacuum state (in absence of the rotation)
$\sigma^{\prime}=\sigma -v$, $\vec{\pi}^{\,\prime}=\vec{\pi}$, $N^{\,\prime}=N$.
 With the symmetry breaking term ${\cal{L}}_{\rm s.b.}^{(2)}$ we have $m_N=gv$, with ${\cal{L}}_{\rm s.b.}^{(1)}$ the nucleon mass term is  given by $m_N=g(v+{\epsilon}/{(2\lambda v^2)})\,.$ The value $v$ is found from the condition of the partial conservation of the axial current, $v=f_\pi\simeq 93$\,MeV, where $f_\pi$ is the pion weak decay constant.
 Taking $m_N\simeq 934$ MeV one recovers the value of the coupling constant $g\simeq 10$.

\section{Chiral fields in rotating systems}\label{chir}
\subsection{Complex $\sigma\pi$ fields in rotating reference frame}
Let us now study behavior of the  chiral vacuum (minimum of energy (\ref{EnerPhipi}) corresponds to  $\mu_{\pi,\Phi}=0$) and a sigma-pion gas at $T=0$ with a dynamically fixed particle number (for $\mu_{\pi}\neq 0$ or $\mu_{\Phi}\neq 0$)  described within  the linear $\sigma$-model in the rigidly rotating cylindrical system at the constant rotation frequency $\vec{\Omega}\parallel z$.
We seek the solution of Eq. (\ref{piSigma}) in cylindrical system of coordinates $(r,\theta, z)$, $\nabla =(\partial_r, \partial_\theta/r, \partial_z)$, , $r=\sqrt{x^2+y^2}$. The coordinate transformation between the laboratory $(t',\vec{r}{\,'})$ frame and the rotating  $(t,\vec{r})$ frame is as follows:  $t'=t$, $x'=x\cos (\Omega t)-y\sin (\Omega t)$, $y'=x\sin (\Omega t) +y\cos (\Omega t)$, $z'=z$. Employing it and that $(ds')^2=\delta_{\mu\nu}dx^{\prime\mu} dx^{\prime\nu}=(ds)^2$, where  $\delta_{\mu\nu}=\mbox{diag}(1,-1,-1,-1)$,  we recover
 expression for the interval in the rotating frame
  \be (ds)^2=(1-\Omega^2 r^2)(dt)^2 +2\Omega y dx dt  -2\Omega x dy dt -(dr_3)^2\,,\nonumber
  \ee
 $r_3= \sqrt{r^2 +z^2}$.   As we see from here, a uniformly rotating system must be finite, othervise  the causality condition $\Omega r<1$ is not fulfilled. Formally requirement of finiteness of the system can be satisfied by imposing a boundary condition at some $r=R$. The latter condition will be discussed  below. So, the rotating frame is determined as \cite{Chen2015}: $e_0^t=e_1^x=e_2^y=e_3^z=1$, $e_0^x=y\Omega$, $e_0^y=-x\Omega$,
 $e_\alpha =e_\alpha^\mu\partial_\mu$, $e_0=\partial_t +y\Omega\partial_x -x\Omega\partial_y$, $e_i =\partial_i$. Lattin index $i=1,2,3$, Greek index $\mu =0,1,2,3$.
Thus in the rotating frame we should  perform  the replacement
\be\partial_t\to \partial_t +y\Omega \partial_x -x\Omega \partial_y=\partial_t -i\Omega\hat{l}_z=
\partial_t -\Omega \partial_\theta\,.\label{shiftrotelectric}\ee

In presence of the gauge fields $A_\mu$ and $\omega_\mu$ in the laboratory frame, the Lagrangian density in the rotating frame renders \cite{Chen2015,Zahed,Guo}:
\begin{eqnarray}
&{\cal{L}}_{\pi}^{\rm rot}=\frac{|(D_t+y\Omega D_x -x\Omega D_y)\Phi_\pi|^2}{2} -\frac{|D_i \Phi_\pi|^2}{2}\nonumber\\
&-
\frac{m^{*2}_\pi |\Phi_\pi|^2}{2}-\frac{\lambda |\Phi_\pi|^4}{4}\,,\label{LVrot}
 \end{eqnarray}
where $D_\mu =\partial_\mu +ieA^{\rm rot}_\mu+ig\omega^{\rm rot}_\mu$, $eA_\mu^{\rm rot}=A_\nu e^\nu_\mu$.  In case of uniform constant magnetic field $\vec{H}\parallel z$ and electric field $eA_0(r)=V(r)$, $g\omega_0 (r)=0$ we have $eA_\mu^{\rm rot}=(V(r)-eH\Omega r^2/2, eHy/2,-eHx/2,0)$. We continue to employ symmetry breaking term in the form ${\cal{L}}_{\rm s.b.}^{(2)}$.

 At $\Omega =0$ we choose as the vacuum state $\sigma =+ v>0$, $\pi_i=0$ in the whole space.
Let us now show that for $\Omega >\Omega_{c}$, where $\Omega_{c}$ is a critical angular velocity,   at $r<R$,
there may appear a phase transition from the state $\sigma = v$ and $\pi_i=0$ to a chiral-vortex state either characterized by  the charged pion condensate  $\Phi_\pi\neq 0$ and $\sigma = v$, $\pi_3=0$ within the model 1, or by the  mean field   $\Phi_\sigma\neq 0$ at $\Phi_\pi= 0$ in the model 2.

Within the { model 1} ($\Phi_\pi\neq 0$, $\sigma=v$)
we seek solution of the equation of motion in the form of the individual vortex \cite{LL9,Tilly-Tilly}:
\be \Phi_\pi =\Phi_{0\pi}\chi (r)e^{i\xi (\theta)-i\mu_\pi t+ip_z z}\,,\quad \Phi_\sigma =v\,,\label{Phifieldformpi}
\ee
with $\Phi_{0\pi}=const$, $p_z =const$. Below we disregard a trivial dependence  $\Phi\propto e^{ip_z z}$ on $z$, since  we will be interested in description of the minimal energy configurations corresponding to $p_z=0$.

Circulation of the $\xi(\theta)$-field yields
\be\oint d\vec{l}\nabla \xi =2\pi \nu\,,\label{circul}\ee
 at the integer values of the winding number $\nu =0,\pm 1,...$ and $\nabla \xi =\nu /r$, $\xi =\nu\theta$,  thereby.
For the case $V=V(r)$, $\vec{A}=0$, $\vec{\omega}=0$, $\omega_0=\omega_0(r)$ of our main interest    the Lagrangian density with taking into account the rotation in the model 1 can be presented as
\begin{eqnarray}
&{\cal{L}}_{\pi}^{V} =-\frac{|\partial_i\Phi_{\pi}|^2}{2}+ \frac{|\widetilde{\mu}\Phi_{\pi}|^2 }{2}-\frac{\lambda |\Phi_{\pi}|^4}{4}
-\frac{m^{*\,2}_\pi |\Phi_{\pi}|^2}{2}\nonumber\\
&+\frac{(\nabla V)^2}{8\pi e^2}+n_p V
\,,\quad \widetilde{\mu}=\mu_\pi  +\Omega\nu-V(r)-g\omega_0\,,\label{LphiOmpiV} \end{eqnarray}
cf. Eqs. (\ref{phiV}),  (\ref{LVrot}).
Equation of motion  for the condensate field is simplified as
\begin{eqnarray}
    [\widetilde{\mu}^2 +\Delta_r -\nu^2/r^2 -m_\pi^{*\,2}] \chi (r) -\lambda |\Phi_{0\pi}|^2\chi^3 (r)   =0\,,\label{piSigma1111pi}
  \end{eqnarray}
cf. Eq. (\ref{piSigma1}),  $\Delta_r=\partial_r^2 +{\partial_r}/{r}$.

Within the { model 2} ($\Phi_\sigma\neq 0$, $\Phi_\pi=0$, $V=0$)
a complex field $\Phi_\sigma$ is affected by the rotation, in spite of the  $\sigma$ and $\pi^0$ separately, being described by the one-component fields,  which are not influenced by the rotation.
We  seek solution of the equations of motion in the form \cite{LL9,Tilly-Tilly}:
\begin{eqnarray} &\Phi_\sigma =\Phi_{0\sigma}\chi (r)e^{i\xi (\theta)-i{\mu}_\Phi t},\, \Phi_\pi =0,\,
%%%%\nonumber\\&\bar{\mu}_\Phi =\mu_\phi -g\omega_0,
\label{Phifieldform}
\end{eqnarray}
$\Phi_{0\sigma}=const$.
The Lagrangian density  renders
\begin{eqnarray}
&{\cal{L}}_\Phi =-\frac{|\partial_i\Phi_{\sigma}|^2}{2}+ \frac{|({\mu}_\Phi +\Omega \nu -g\omega_0)\Phi_{\sigma}|^2 }{2}\nonumber\\
&-\frac{\lambda (|\Phi_{\sigma}|^2 -v^2)^2}{4}
-\frac{m^{*\,2}_\pi \pi_3^2}{2}
\,.\label{LphiOm} \end{eqnarray}

Neglecting rapidly oscillating terms we may replace $\sin^2 (\nu\theta -{\mu}_\Phi t)$ by $1/2$,
  cf. Eq. (\ref{piSigma111}).
Then equation of motion  for the stationary field reads
\begin{eqnarray}
&[\widetilde{\mu}^2 +\Delta_r -m_\pi^{*\,2}/2 ]\chi (r)  \nonumber\\&+\lambda [v^2 -|\Phi_{0\sigma}|^2\chi^2(r)]\chi(r)    =0\,,\label{piSigma1111}
  \end{eqnarray}
 For the model 2 we have $\widetilde{\mu}=\bar{\mu}_\Phi =\mu_\Phi +\Omega\nu -g\omega_0$.

 The angular momentum associated with the boson field $\Phi_{\pi,\sigma}$  is given by
\begin{eqnarray}&\vec{L}_{\pi,\Phi}=\int d^3 X [ \vec{r}\times\vec{P}]\,\,, \nonumber\\
&\vec{P}=\frac{1}{2}[\frac{\partial{\cal{L}}_{\pi,\Phi}}{\partial\partial_t{\Phi}_{\pi,\sigma}}\nabla \Phi_{\pi,\sigma} +\frac{\partial{\cal{L}}_{\pi,\Phi}}{\partial\partial_t{\Phi}^*_{\pi,\sigma}}\nabla \Phi^*_{\pi,\sigma}],\,\,\nonumber\\&P_\theta =\widetilde{\mu}  \Phi^*_{\pi,\sigma}\nabla \Phi_{\pi,\sigma}/i \,.\label{pphi}\end{eqnarray}
In case of the vortex (index $v$) placed in the center of the coordinate system $P^v_\theta = \widetilde{\mu} |\Phi_{\pi,\sigma}|^2\nu/r =\rho_{\pi,\Phi} \nu /r$ and $\vec{L}^v =\int d^3 X [ \vec{r}\times\vec{P}^v]$.
Thus we obtain
\be
L^v_z =  2\pi d_z \int_0^R r dr  \nu\rho_{\pi,\Phi}
%v^2 m_{\Phi'}
\,,\label{Lvortsingle1}\ee
where the charged particle/field density is given by \be\rho_{\pi,\Phi}=\frac{\partial{\cal{L}}^V_{\pi,\Phi}}{\partial\mu_{\pi,\Phi}}
=\widetilde{\mu}|\Phi_{\pi,\Phi}|^2\,\label{densmod1rot}.\ee
In case $\rho_{\pi,\Phi}\simeq const$ we would have $L_z^v \simeq \nu \pi d_z {R}^2 \rho_{\pi,\Phi}$.
If the  vortex is placed at a distance $b$ from the center, one should replace ${R}$ in (\ref{Lvortsingle1}) to $\sqrt{{R}^2-b^2}$, cf. \cite{PethickSmith}. Note that the same Eq. (\ref{Lvortsingle1}) follows from Eqs. (\ref{LphiOmpiV}), (\ref{LphiOm}) after usage  that the angular momentum is $\vec{L}_{\pi,\Phi} =\int d^3 r \partial {\cal{L}}^V_{\pi,\Phi}/\partial \vec{\Omega}$. For $\nu\gg 1$ we will name the resulting vortex field the supervortex. The superfluid as a whole  can either  remain at rest or it can participate in a rotation. In the latter case the superfluid may %rotate as
mimic a rigid body rotation with the angular velocity $\omega$ close to $\Omega$.
The total boson angular momentum is the sum of the vortex term and the term related to the rotation of the superfluid with the angular velocity $\omega$. First we will focus on the case when $\omega =0$. The case $\omega\neq 0$ is relevant for the rotation of the  condensed Bose gas in some frequency interval when the vortices form a lattice.  It  will be discussed in subsection \ref{lattice} and in Appendix 2.

 Above we considered the case when the rotation frequency $\Omega$ of the rotating frame is fixed. The vortex field arises when its  energy in the rotation reference frame becomes negative.

\subsection{  Rotation- and laboratory frames. Rigid-body rotation}

The question how to treat the rotating reference frame and the response  of the Bose field vacuum and the Bose gas at zero temperature (i.e. the Bose-Einstein condensate) on the rotation in this frame is rather subtle owing to necessity to fulfill the causality condition $r<1/\Omega$.  We may associate the rotation frame with a rotating rigid body of a finite transversal size.
For instance, we may  consider either the vacuum or the pion-sigma gas  inside a  long empty cylindrical vessel of a large internal transversal radius $R$, external radius $R_{>}$, hight $d_z\gg R$,  mass $M$  and constant mass-density $\rho_M$, rotating in the $z$ direction with constant cyclic frequency $\Omega$ at  $\Omega <\Omega_{\rm caus}=1/R_{>}$, as the requirement of causality.  Limiting cases $R\to 0$ and $R_{>}\to R+0$ as well as $R_{>}\gg R$ are allowed.   For the Bose gas described by a complex field there are two possibilities: (1)  it  responses on the rotation of the vessel  creating the  vortices, (2)  it does not rotate, cf. \cite{LL9,Tilly-Tilly,PethickSmith,PitString}. The latter possibility is realized for rotation frequencies $\Omega <\Omega_{c1}$, where $\Omega_{c1}$ is a critical angular velocity. For the vacuum (the ground state in absence of particles) we also should study two possibilities: (1) in the rotation reference frame it responses on the rotation of the vessel  producing a vortex field and (2)   it does not rotate, remaining the same as in the laboratory reference frame.

Further, there are  two possibilities: (i) conserving rotation frequency $\vec{\Omega}_{fin}=\vec{\Omega}_{in}$ of the rotating rigid body, and (ii) conserving angular momentum $\vec{L}_{fin}=\vec{L}_{in}$.

In case (i) the kinetic energy of the vessel at the nonrelativistic rotation   measured in the laboratory (resting) reference frame,
\be{\cal{E}}_{in}=\pi \rho_M \Omega_{in}^2 d_z (R_{>}^4 -R^4)/4
%+{\cal{E}}_{\pi,\Phi}[\Omega=0]
\,,\label{EpsilonM}\ee
does not change with time.
The loss of the energy due to a radiation  is recovered    from  an external source. We will consider the situation when  in the rest frame the pion field  condensate does not appear from  the vacuum even in presence of external fields. It is so at least provided external fields are not too strong. However we still should consider a possibility for formation of the pion condensate from the vacuum in the rotation reference frame. In presence of  Bose excitations the final energy of the system is ${\cal{E}}_{ fin}={\cal{E}}_{in}+{\cal{E}}_{\pi,\Phi}[\Omega_{in}]$, ${\cal{E}}_{\pi,\Phi}[\Omega_{in}]$ is the rotation part of the energy associated with a Bose condensate in the  rotating system.
 The condition for the formation of the condensate in the rotating piece $r<R$ of the vacuum  is as follows
 \be{\cal{E}}_{\pi,\Phi}[\Omega_{in}]
 %={\cal{E}}_{\pi,\Phi}[\Omega_{in} ]-{\cal{E}}_{\pi,\Phi}[\Omega=0]
 <0\,.\label{standEin-finRot}\ee
The same condition holds, if we deal with the gas with fixed particle number, with the only difference that for the gas $\mu_{\pi,\Phi}\neq 0$, and the latter quantity is determined from the condition of the fixed particle number.

 In case (ii) the vessel is rotated owing to the  initially applied angular momentum $\vec{L}_{in}=\int d^3 X [ \vec{r}_3\times\vec{P}_{in}]$, $\vec{P}_{in}=\rho_M[\vec{\Omega}_{in},\vec{r}_3]$,\,
 %${r}_3=\sqrt{r^2+z^2} $,
  \be \vec{L}_{in}=I_{in}\vec{\Omega}_{in}=\pi \rho_M \vec{\Omega}_{in} d_z(R^4_{>}-R^4)/2\,,\label{Lin}\ee
  which value is conserved and can be redistributed between the vessel and the condensate of the chiral field, if the formation of the condensate is energetically favorable. $I_{in}$ is the moment of inertia. We have
  \begin{eqnarray}
  &\vec{L}_{in}=\vec{L}_{M,fin}+\vec{L}_{\pi,\Phi}^{\rm lab}\,,\, \nonumber\\
  &\vec{L}_{M,fin}=\pi \rho_M \vec{\Omega}_{fin} d_z(R^4_{>}-R^4)/2\,.\end{eqnarray}
A somewhat similar problem has been studied in \cite{V93jetp,Vexp95,Kolomeitsev:2016isb,Voskresensky:2022fzk}  in case of rectilinear motion of the wall in the superfluid.
 The final energy  is
 \be{\cal{E}}_{fin}=\pi \rho_M \Omega_{fin}^2 d_z (R_{>}^4 -R^4)/4 +{\cal{E}}_{\pi,\Phi}^{\rm lab}\,, \ee
and for the gas with fixed particle number
  \be{\cal{E}}_{\pi,\Phi}^{\rm lab}={\cal{E}}_{\pi,\Phi}[\nu, \Omega =0]\,.\ee
From the latter equation in case (ii) of the conserving initial angular momentum we obtain
\begin{eqnarray}
&{\cal{E}}_{fin}= {\cal{E}}_{in}-\vec{L}_{\pi,\Phi}^{\rm lab}\vec{\Omega}_{in}+ {\cal{E}}_{\pi,\Phi}^{\rm lab}[\nu, \Omega =0]\nonumber\\&+(L_{\pi,\Phi}^{\rm lab})^2/[\pi \rho_M  d_z(R^4_{>}-R^4)]\,.\label{caseii}\end{eqnarray}
In case of a weak condensate field (for $L_{\pi,\Phi}^{\rm lab}
%=\nu N_{\pi,\Phi}
\ll L_{in}$), which we will be interested in, the last term can be neglected and
\be \delta {\cal{E}}={\cal{E}}_{fin}- {\cal{E}}_{in}\simeq -L_{\pi,\Phi}^{\rm lab}\Omega_{in}+ {\cal{E}}_{\pi,\Phi}^{\rm lab}[\nu, \Omega =0]\,.\label{standEin-fin}\ee
The vortex-condensate field appears provided $\delta {\cal{E}}<0$. Below we will show that for the description of the nonrelativistic pion-sigma  gas at a fixed particle number conditions (\ref{standEin-finRot}) and (\ref{standEin-fin})  coincide. On the other hand the vacuum at $\Omega =0$ is described by $\sigma^2 =v^2$ but $\Phi_{\pi,\sigma}=0$
and the condition (\ref{standEin-fin}) becomes helpless. The response of  the Bose vacuum  on the rotation is manifested in the rotating frame.

Generalization of the expression for the kinetic energy given by the first term Eq. (\ref{EpsilonM}) to the case of a relativistic rotation of the vessel is as follows
\begin{eqnarray}
&{\cal{E}}_{in}^{\rm kin}=\int^{R_{>}}_R 2\pi \frac{\rho_M  d_z r dr}{\sqrt{1-\Omega_{in}^2r^2}}-\pi\rho_M d_z (R_>^2-R^2)\nonumber\\
&=-\frac{2\rho_M\pi d_z}{\Omega_{in}^2}(\sqrt{1-\Omega_{in}^2R_>^2}-\sqrt{1-\Omega_{in}^2R^2})\nonumber\\&-\pi\rho_M d_z (R_>^2-R^2)\,.
 \end{eqnarray}
 In case $R_>\simeq R+\delta R$, $\delta R\ll R$,
 $\delta\Omega=\Omega_{in}-\Omega_{fin}\ll \Omega_{in}$  we have \be{\cal{E}}_{in}^{\rm kin}-{\cal{E}}_{fin}^{\rm kin}\simeq
 \frac{2\rho_M\pi d_z R^2\Omega_{in}R\delta\Omega \delta R}{(1-\Omega_{in}^2R^2_>)^{3/2}}\,. \ee

 Criterion of   causality requires that the condition $\Omega_{in}R_><1$ should be fulfilled at the rigid-body rotation of the vessel. However there is still a possibility to further increase the rotation frequency $\Omega_{in}$, provided  it is energetically favorable  to destroy the supervortex or the lattice of vortices and  redistribute  the angular momentum   between the individual separately rotating vortices. This possibility will be considered in Section \ref{lattice}.

\subsection{Fields in absence of self-interaction}\label{sect-ideal}
\subsubsection{Equation of motion, boundary conditions and   energy in rotation frame}

First let us consider the case $\lambda =0$, which is  actually not specific for the $\sigma$ model.
From Eqs. (\ref{piSigma1111pi}), (\ref{piSigma1111})  we  recover   equation of motion
\be
\left[\partial_r^2 +\frac{\partial_r}{r} -\frac{\nu^2}{r^2}+(\widetilde{\mu}^2-\widetilde{m}^2)
%\nu^2\Omega^2-\frac{m^{*2}}{2}
\right]\chi(r)=0\,, \label{KGF}
\ee
 where we introduced the quantities $\widetilde{m}^2=m^{*\,2}_\pi$ in case of the model 1 and  $\widetilde{m}^2=m^{*\,2}_\pi/2$ for model 2 provided we continue to use the symmetry breaking term in the form ${\cal{L}}_{\rm s.b.}^{(2)}$, $\widetilde{\mu}$ is determined by Eq. (\ref{LphiOmpiV}).
   Equation (\ref{KGF}) describes a spin-less relativistic particle of the energy $\epsilon_{n,\nu}=\mu_{\pi,\Phi}$,  mass $\widetilde{m}$ and $z$ projection of the angular momentum $L_z=\nu$, placed  in the  axially symmetric  potential well with the zero-component of the vector potential  $U(r)=- \Omega\nu +V(r)+g\omega_0$  for $r<R$, with $V\neq 0$ in the model 1.  Behavior at $r>R$ depends on the  boundary condition put at $r=R$.
   In case of the $\sigma +i\pi_3$ field (within the model 2) one should put $V=0$ at all $r$.

 Now we are interested in the description of the ground state, then ${\mu}_{\pi,\Phi}=\mbox{min}\{\epsilon_{n,\nu}\}$ plays a role  of the chemical potential. The term $\Omega\nu{\mu}_{\pi,\Phi} |\Phi_{\pi,\Phi}|^2$ in the energy density can be associated with the Coriolis force and the term $\Omega^2\nu^2 |\Phi_{\pi,\Phi}|^2/2$, being  an attractive relativistic  $\propto 1/c^2$ contribution to the centrifugal force term  $\nu^2/r^2$.  The
 Schr\"odinger equation follows from the Klein-Gordon equation (\ref{KGF}) after doing the replacement ${\mu}_{\pi,\Phi}\to \widetilde{m}+\mu_{\rm n.r.}$ and subsequent dropping of  small quadratic terms $(\mu_{\rm n.r.}+\Omega\nu -V-g\omega_0)^2$. Then Eq. (\ref{KGF}) acquires the form
 \be \left[-\frac{\Delta_r}{2\widetilde{m}} -\Omega\nu+V(r) +g\omega_0+\frac{\nu^2}{2\widetilde{m}r^2}\right]\chi =E_{\rm n.r.}\chi,\label{Schrod}\ee
  $U_{ef}=-\Omega\nu +V(r)+g\omega_0  +{\nu^2}/(2\widetilde{m}r^2)$, $E_{\rm n.r.}=\mu_{\rm n.r.}$. As we see, the rotation in the cylindrical rotating frame acts similarly  to a constant (and attractive for $\Omega\nu>0$) electric potential acting on a nonrelativistic particle with the angular momentum $\nu$.    Treating the vessel  as the potential box  with infinite walls, we may use the boundary  condition $\chi (r=R)=0$.

 The Schr\"odinger equation should not change under simultaneous replacements $\vec{r}_3\to \vec{r}^{\,\prime}_3+\vec{W}t^{\prime}$ and $t=t^{\prime}\,$,\,
  provided
 \be
 \Psi (t,\vec{r}_3)=e^{i\chi (t,\vec{r}_3)} \Psi^{\prime}(t^{\prime},\vec{r}_3^{\,\prime})\,,\label{psitheta}\ee
  where $\chi$ is a real function, $\vec{W}$ is a constant vector. Indeed, the latter transformation does not modify probability density, i.e. $|\Psi (t,\vec{r}_3)|^2=|\Psi^{\prime}(t^{\prime},\vec{r}^{\,\prime}_3)|^2$.
 Then using
 \be
 \partial_{\vec{r}^{\,\prime}_3}=\partial_{\vec{r}_3}\,, \quad \partial_{t^{\prime}}=\partial_t +\vec{W}\partial_{\vec{r}_3}\label{Galpsi}\ee
 we find
 \be
 \chi (t,\vec{r}_3)={\widetilde{m}\vec{W}\vec{r}_3}-\frac{\widetilde{m}W^2 t}{2}\label{theta}
   \ee
apart irrelevant constant. Using these expressions we see that in the nonrelativistic case the rotation can be  introduced in the Schr\"odinger equation with the help of the replacement, cf. \cite{FischerBaym2003},
\be -\frac{\Delta }{2\widetilde{m}}\to -\frac{(\nabla -i \widetilde{m}\vec{W})^2}{2\widetilde{m}}-\frac{\widetilde{m}\vec{W}^2}{2}\,
\label{rotSchrod}\ee
with $\vec{W}=[\vec{\Omega}\times \vec{r}_3]$ that results in the same Eq. (\ref{Schrod}).
This equivalence exists only for a nonrelativistic rotation.

Thus we see that uniform rotation can be introduced in nonrelativistic case  similarly to  a uniform rather weak magnetic field described by the vector-potential $\vec{A}=\frac{1}{2}[\vec{H},  \vec{r}_3]$. In the general relativistic case the shift of variables $\partial_t\to \partial_t-\Omega \partial_\theta$ in the Klein-Godron equation done in the rotation frame is not equivalent to the shift of the  spatial variables $\nabla\to \nabla -i\widetilde{m}[\vec{\Omega},\vec{r}_3]$ in the Hamiltonian and  subtraction of the $\frac{\widetilde{m}\vec{W}^2}{2}$ term associated with the motion of the system as a whole, cf.  \cite{Chen2015}.

Further,  dealing with the model 1 let us for simplicity consider the case $V\simeq -V_0=const$. This approximately constant value can be treated as a contribution to the chemical potential.
For example we may assume that  an ideal rotating vessel is placed inside the cylindrical co-axial capacitor. Then  in absence of the classical boson field the electric potential $V(r)=-V_0=const$ at distances  $r<R$ of our interest. Appearance of the field $\Phi_\pi\neq 0$ produces a dependence of $V$ on $r$. However, if  $\Phi_\pi$ is a rather small, we can continue to consider $V= -V_0\simeq const$. In case of the model 2 the electric field and the boson field decouple, as we have mentioned.

Employing dimensionless variable $x=r/r_0^{\rm lin}$, with
\be r_0^{\rm lin}=1/\sqrt{\bar{\mu}^2-\widetilde{m}^2}\,, \, \bar{\mu}=\mu_{\pi,\Phi}+\Omega\nu+V_0-g\omega_0\,,
\label{rlin}\ee
for $\bar{\mu}> \widetilde{m}$,  $V_0=const$ from Eq. (\ref{KGF}) we obtain equation
 \be (\partial^2_x +x^{-1}\partial_x -\nu^2/x^2)\chi+\chi =0\,.\label{filvorteqdim1}
 \ee

Simplest appropriate boundary conditions are
\be\chi (0)=0\,,\quad \chi (R/r_0^{\rm lin})=0\,.\label{boundcon}\ee
The latter condition is equivalent to the existence  of the  infinite wall at $r=R$ in the single-particle quantum mechanical problem.

Further to be specific let us consider $\Omega, \nu>0$.
 Appropriate solution of Eq. (\ref{filvorteqdim1})  is the Bessel function
 \be\chi (r)=J_\nu (r/r_0^{\rm lin})\label{Bessel}\ee
  for $\nu >0$,
  cf. \cite{Gradshtein}. For $x\to 0$ we have $J_\nu \sim x^\nu$. The energy of the $n=1$ level is determined by the first zero of the function  $J_\nu (R/r_0^{\rm lin})=0$, $j_{n=1,\nu =0}\simeq 2.403$. The  $n=1,\nu=1$ zero yields  $j_{1,1}=R/r_0^{\rm lin}\simeq 3.832$,  $j_{1,\nu}$ increases with increase of integer values of $\nu$.
   %and $j_{1,\nu\gg 1}\simeq \nu$.
   For $x\gg \nu$ we have $J_\nu (x)\simeq \sqrt{2/(\pi x)}\cos (x-\pi\nu/2-\pi/4)$.
From here we find approximate asymptotic value $j^{\rm as}_{n,\nu}\simeq \pi(\nu +1/2+2n-1)/2$, where the integer number
$n\geq 1$ is the corresponding zero of the function $\cos (x-\pi\nu/2 -\pi/4)$, and thereby $j_{1,1}^{\rm as}\simeq 5\pi/4\simeq 3.927$ that only slightly differs from the exact solution $3.832$.
%$j_{1,\nu}\simeq \pi(2\nu +3)/4$ and $j_{1,\nu =1}\simeq 3.927$.
For $\nu\gg 1$ we have $j_{1,\nu}^{\rm as}\to \nu+1.85575\nu^{1/3}$, e.g., $j_{1, 100}\simeq 108.84$, $j_{1,10^4}\simeq 10040$.

Equation of motion (\ref{filvorteqdim1}) together with the boundary conditions (\ref{boundcon}) is satisfied only for discrete values of the rotation frequency.
 Employing the boundary condition $\chi(x=R/r_0^{\rm lin}=j_{n,\nu})=0$  for the energy level with the quantum numbers $n,\nu$  we find
 \begin{eqnarray} &\epsilon_{n,\nu}=\mu_{\pi,\Phi}=-\Omega\nu -V_0 +g\omega_0
 \nonumber\\&+\widetilde{m}\sqrt{1+j_{n,\nu}^2/(R^2 \widetilde{m}^2)}\,,
 \label{grlev}\end{eqnarray}
 $\bar{\mu}=\sqrt{\widetilde{m}^2+j_{n,\nu}^2/R^2}.$
It is important to notice that with increase of the quantity $\Omega\nu +V_0-g\omega_0$
the $n,\nu\neq 0$ levels become more bound than the level $n=1,\nu=0$. For example
for $g\omega_0=0$, $R\widetilde{m}\gg 1$ the level $n=1,\nu=1$ crosses the level $n=1,\nu=0$ at $\Omega R\simeq
%-V_0 R+
2.41/(R \widetilde{m})$.
For such heavy nuclei as $U$ the nucleus radius is $R\simeq 7$fm$\simeq 5/m_\pi$, $V_0 R\simeq 1.2 Ze^2\simeq 0.8$ and following our estimate the crossing of the levels for charged pions may occur only for $\Omega R\gsim 0.5$,  and for the neutral $\sigma - \pi$ excitations for $\Omega R>0.7$.  Similar estimates hold for the nuclear fireball prepared in peripheral the heavy-ion collisions.
From (\ref{grlev}) we  also find that the roots  $\epsilon_{n,\nu}$ for $V_0=g\omega_0=0$ do not reach zero for
 $\Omega R<1$. Thus   for $V_0=g\omega_0=\lambda=0$ the fields $\Phi_{\pi,\sigma}$ would not appear at the rotation of the vacuum.

 Using Eqs. (\ref{LphiOmpiV}), (\ref{LphiOm}) and boundary conditions $(\Phi\Phi^{\prime}_r )_{r=R}= (\Phi\Phi^{\prime}_r )_{r=0}$, taking  $n_p=const$, $n_p 4\pi R^3/3=Z$, $V=-V_0\simeq const$ in case  $\lambda =0$, which we now consider, we recover the energy:
\begin{eqnarray}
&{\cal{E}}_{\pi,\Phi}(\Omega) ={\cal{E}}_{\pi,\Phi}^{\rm l,0} +{\Phi_{0\pi,\sigma}^2d_z} \pi\int_0^R r dr\chi^2(r)\mu_{\pi,\Phi}^2+V_0 Z
\nonumber\\
%%&={\cal{E}}_{\pi,\Phi}^{\rm l} +
%%{\Phi_{0\pi,\sigma}^2d_z} 2\pi\int_0^R r dr\chi^2(r)\mu_{\pi,\Phi}\widetilde{\mu}+V_0Z
&={\cal{E}}_{\pi,\Phi}^{\rm l} +\mu_{\pi,\Phi}N_{\pi,\Phi}+V_0 Z\nonumber\\
&={\cal{E}}_{\pi,\Phi}^{\rm l} +
{\Phi_{0\pi,\sigma}^2d_z} 2\pi\int_0^R r dr\chi^2(r)\widetilde{\mu}^2 \nonumber\\
&-L^v\Omega
-V_0 (N_\pi -Z)+g\omega_0N_{\pi,\Phi}\,,\label{Eg0}
\end{eqnarray}
where $L^v=\nu N_{\pi,\Phi}\,,$
%{\Phi_{0\pi,\sigma}^2d_z} 2\pi\int_0^R r dr\chi^2(r)\widetilde{\mu}$,
$N_{\pi,\Phi} = {\Phi_{0\pi,\sigma}^2d_z} 2\pi\int_0^R r dr\chi^2(r)\widetilde{\mu}$,
\begin{eqnarray}&{\cal{E}}_{\pi,\Phi}^{\rm l,0}={\Phi_{0\pi,\sigma}^2d_z} \pi\int_0^R r dr\chi(r)\nonumber\\&\times
\left[-\partial_r^2 -\frac{\partial_r}{r} +\frac{\nu^2}{r^2}-[(\Omega\nu+V_0-g\omega_0)^2-\widetilde{m}^2]
%\nu^2\Omega^2-\frac{m^{*2}}{2}
\right]\chi(r)\,,\nonumber\\
&{\cal{E}}_{\pi,\Phi}^{\rm l}={\Phi_{0\pi,\sigma}^2d_z}\pi\int_0^R r dr\chi(r)\nonumber\\&\left[-\partial_r^2 -\frac{\partial_r}{r} +\frac{\nu^2}{r^2}-(\widetilde{\mu}^2-\widetilde{m}^2)
%\nu^2\Omega^2-\frac{m^{*2}}{2}
\right]\chi(r)\,,\label{Ekinvac}\end{eqnarray}
$N_{\pi,\Phi}$ is the number of particles placed on the energy level $\epsilon_{n,\nu}$, $\Phi_{0\pi,\sigma}$ is an arbitrary constant.
   To get Eq. (\ref{Eg0}) we used equation of motion (\ref{KGF})  and boundary conditions $(\Phi\Phi^{\prime}_r )_{r=R}= (\Phi\Phi^{\prime}_r )_{r=0}$. Within the model 1 due to  the  charge neutrality condition inside the system of a sufficiently large size we have $N_\pi =Z$.
   To consider general case  $V\neq const$ one should add to the energy density the term $-(\nabla V)^2/(8\pi e^2)$. After taking into account the Poisson equation for $V$ it results in that in the first line (\ref{Eg0}) one should replace $V_0 Z\to -\int_0^R (n_p+\widetilde{\mu} |\Phi|^2)\pi r dr V d_z$.

We could  employ the boundary condition
 $\chi^{\prime}(r=R)=0$ instead of the  condition $\chi(r=R)=0$ that we have used. In both cases there is no current through the surface $r=R$. Thereby such a change of the boundary condition would not affect our conclusion that at $V_0=g\omega_0=0$ the energy level does not cross zero. Also we note that in cases of the vacuum and the Bose condensate in the vessel usage of one of mentioned boundary  conditions is motivated provided the typical frequency of atomic transitions in the solid wall, $\omega_{\rm at}$, is larger than difference between energies of  the first excited and ground state energy levels, $\sim (1+\nu)^2/(R^2 \widetilde{m})$. The latter condition is well  satisfied for  $R>a$, where $a$ is the typical atomic size.

Setting solution (\ref{grlev}) to the second line of Eq. (\ref{Eg0}) and employing that ${\cal{E}}_{\pi,\Phi}^{\rm l}=0$ on the solutions of equation of motion  (\ref{filvorteqdim1}), we find
\begin{eqnarray}
&{\cal{E}}_{\pi,\Phi} (\Omega)={\Phi_{0\pi,\sigma}^2d_z}\nonumber\\ &\times 2\pi\int_0^R r dr\chi^2(r)\epsilon_{n,\nu} \widetilde{m} \sqrt{1+j_{n,\nu}^2/(R^2 \widetilde{m}^2)}
\,.\label{bosenrot}\end{eqnarray}

\subsubsection{ Empty rotating vessel. A  ``rotating vacuum''}\label{subsubsection-rotves}

{\bf Conditions  of vacuum instability in rotating frame in case (i).}

 %However the latter equation can be satisfied only for $V_0\neq 0$ and/or $g\omega_0\neq 0$, since  we have found that the roots $\epsilon_{1,\nu}$ remain positive  for $V_0=g\omega_0=0$  at the condition $\Omega R<1$.
%%for ${\nu}\ll r_0^{\rm lin}/R= j_{\nu,n}$,
% R/r_{0}^{\rm lin}\sim \widetilde{m}R$,

In QED in electric potential well for the $\pi^-$ in absence of the matter, the $\pi^+\pi^-$ pairs are produced from the vacuum when the $\pi^-$ ground state level crosses the energy $-m_\pi$. The $\pi^+$ go off to infinity.  In a piece of the nuclear matter $\pi^+\pi^-$ pairs can be  produced in reactions $n\leftrightarrow p+\pi^-$ when the level energy reaches zero, cf. \cite{Migdal:1977rn,Migdal1978}.

In case $V_0=g\omega_0=0$ energetically favorable solution in the rotating frame corresponds to $\epsilon_{n,\nu}>0$ and  ${\cal{E}}_{\pi,\Phi}>0$ for $\Phi_{\pi,\sigma} \neq 0$ and thereby $\Phi_{\pi,\sigma} =0$ in the vacuum in both models 1 and 2. The solutions of Eq. (\ref{grlev}) corresponding to $\epsilon_{n,\nu}<0$,  may exist  at the condition $\Omega R<1$ only  for $V_0,g\omega_0\neq 0$. If the chemical potential $\mu_{\pi,\Phi}$ crosses zero, the boson field can be produced from the vacuum. For attractive $V$ and $\nu>0$, $\pi^-$ mesons  occupy the dangerous level $(1,\nu)$, whereas $\pi^+$ can be absorbed on the wall of the vessel. In case of $\lambda =0$ under consideration now, the energy gain  $\propto \Phi_{0\pi,\sigma}^2$ can be made arbitrary large, as it follows from Eq. (\ref{bosenrot}).

 The values of the  rotation frequency, at which the equation of motion (\ref{grlev}) could be fulfilled for $\mu_{\pi,\Phi} =\epsilon_{1,\nu}\leq 0$, $V_0\neq 0$ and/or for $g\omega_0<0$ are given by
\begin{eqnarray} &\Omega\geq \Omega_c=\Omega (\epsilon_{1,\nu}=0)\nonumber\\&=(\widetilde{m}\sqrt{1+j_{1,\nu}^2/(R^2 \widetilde{m}^2)}-V_0+g\omega_0)/\nu\,.\label{critOmvac}\end{eqnarray}

In case $1\leq \nu =c_1 \widetilde{m}R\ll \widetilde{m}R$, i.e. for $c_1\ll 1$,  $\widetilde{m}R\gg 1$ from Eq. (\ref{grlev}) we have
\be
\epsilon_{1,\nu}\simeq -V_0+g\omega_0-\Omega\nu +\widetilde{m}+...\label{largenulimless}\ee
The levels $\epsilon_{1,\nu}$ reach zero only for  $V_0-g\omega_0> \widetilde{m}(1-c_1)$. The critical rotation frequency is then given by
\be\Omega_c =\Omega (\epsilon_{1,\nu}=0, c_1\ll 1)\simeq (\widetilde{m}-V_0+g\omega_0)/\nu >0.\label{Omcrsmall}
\ee
 As we see, the critical rotation frequency  $\Omega_c$ decreases with increasing $\nu$.

In case of the solution describing a supervortex with $\nu=c_1\widetilde{m} R\gg \widetilde{m} R\gg 1$ (for $c_1\gg 1$) from (\ref{grlev}) we have
\begin{eqnarray}
&\epsilon_{1,\nu}\simeq -V_0+g\omega_0+(-\Omega R +1)\nu/R+R\widetilde{m}^2/(2\nu)\nonumber\\&+1.86\nu^{1/3}/R+...\label{largenulim}
\end{eqnarray}
Setting in (\ref{largenulim})  the  limiting value $\Omega^{\rm caus}=1/R$ we see that $\epsilon_{1,\nu}\to -V_0+g\omega_0+\widetilde{m}/(2c_1)+...$. Thus the level $\epsilon_{1,\nu}$ may reach zero for $\Omega R<1$ for $V_0-g\omega_0>\widetilde{m}/(2c_1)$. So, with increase  of the angular momentum of the rotating vessel   the critical value $V_0-g\omega_0$, at which the level $\epsilon_{n=1,\nu}$  reaches zero, decreases.
Formation of the supervortex state becomes energetically favorable  for $c_1\gg 1$ at
\begin{eqnarray}
&\Omega>\Omega_c =\Omega (\epsilon_{1,\nu}=0,c_1\gg 1)\simeq \frac{1}{R}-\frac{V_0-g\omega_0-\widetilde{m}/(2c_1)}{c_1\widetilde{m}R}\,,\nonumber\\
&V_0-g\omega_0>V_{0c}-g\omega_{0c}=\widetilde{m}/(2c_1)\,. \label{condVom}\end{eqnarray}
The larger is $\nu$ the smaller is the value $V_{0c}-g\omega_{0c}>0$. In case (i) the critical value $V_{0c}-g\omega_{0c}$ tends to zero whereas in case (ii) the value $c_1$ is fixed by the conservation of the initial angular momentum. In case of the charged field (in the model 1) we should still care of the charge conservation.

{\bf Examples of instability of  rotated vacuum.}  Let us  give some examples when the rotated vacuum is unstable to formation of the vortex pion-sigma field.

 (a) {\em Rotating vessel inside a  charged capacitor.}  Let  $n_p=0$. In case when an ideal  rotating vessel is placed inside a cylindrical co-axial charged capacitor (with cylindrical plates placed at $r=R_{ex}$ and $r=R_{in}$ for $R_{ex}>R_{in}>R_{>}$) we have $Z_{in} +2\pi  d_z\int_0^R r dr\rho_\pi (r) =-Z_{ex}$, where $Z_{in}$ is the charge placed on the internal surface of the capacitor and $-Z_{ex}$ is the charge placed on the external surface. The strength of the electric field  between the plates is $E(r>R_{in})=-Z_{ex} /( 2\pi r d_z)$ and $V(r=R_{in})=R_{in} E(R_{in})\ln (R_{ex}/R_{in})$. For $\rho_\pi d_z \pi R^2\ll Z_{ex}$ we have $V(r<R)\simeq V(r=R_{in})$.
As we have estimated above, the conditions  $|V|> |V_c|\simeq \widetilde{m}/(2c_1)$ and even $|V|\gsim m_\pi$ can be easily fulfilled in both (i) and (ii) cases.

(b) {\em Redistribution of charge inside the rotating vessel.} Let us rise a question whether  the vortex field  can be produced  in absence of the capacitor. In case of the rotation of the electrically neutral  empty vessel,  the charge of the supervortex, if the latter is  formed in the center of the vessel, should be  compensated by the oppositely charged particles (antiparticles)  shifted closer to the inner surface of the rotating vessel such that $\int_0^R\rho_\pi r dr =0$.
We recall that  $j_{1,1}=r_0^{\rm lin}/R\simeq 0.26$ and $j_{1,\nu\gg 1}\simeq 1/\nu$ and thereby  $r_0^{\rm lin}\ll R$ in all cases we are interested in. We have $2\pi r E(r)\simeq 4\pi |e|\int_0^r \rho_\pi 2\pi r dr$. Let for simplicity  $g\omega_0=0$.
  Assuming for a rough estimation for $\nu\gg 1$ that $J^2_\nu (r/r_0^{\rm lin})\simeq 2 r_0^{\rm lin}\cos^2 (r/r_0^{\rm lin}-\pi\nu/2-\pi/4)/(\pi r)$, using Eq. (\ref{densmod1rot}) we get $E\simeq  4|e|\bar{\mu}\Phi^2_{0\pi} r_0^{\rm lin}$ and
  $V_0(r\sim R)\sim  4e^2\bar{\mu}\Phi^2_{0\pi} r_0^{\rm lin}R$. The level energy $\epsilon_{1,\nu}$ reaches zero for $8 e^2 \bar{\mu}\Phi^2_{0\pi} r_0^{\rm lin}R \gsim \widetilde{m}/c_1$. Since here $\Phi^2_{0\pi}$ is an arbitrary constant,  this condition can be  easily satisfied for both cases (i) and (ii).
 In case $V\neq const$ the electric field produces extra positive energy term $\int_0^R(E^2/8\pi)2\pi rdr d_z\simeq 2 e^2\bar{\mu}^2\Phi_0^4(r_0^{\rm lin})^2R^2d_z$, which proves to be smaller than the negative contribution (\ref{bosenrot}) at least for $8e^2\bar{\mu}\Phi^2_{0\pi} r_0^{\rm lin}R \gg \widetilde{m}/c_1$.
 The uniform rotation acting as an effective electric field  provides  separation of the electric charges and formation of an electric potential, which can be sufficient for   production of  a supervortex of the  charged pion field.
Thereby  statement that cold vacuum in absence of external fields cannot rotate can be questioned.

(c) {\em Pion-sigma field in peripheral heavy-ion collisions.}
In heavy-ion collisions at LHC conditions the typical parameters of the pion fireball estimated in the resonance gas model \cite{Stachel} are as follows: the temperature $T\simeq 155$ MeV, the volume is $5300$fm$^{3}$, the $\pi^{\pm}$ density is $\rho_{\pi}\simeq m^3_\pi$ and  we estimate the  electric potential as  $V_0\sim Ze^2/R\sim  0.2 m_\pi$ for central collisions. For peripheral collisions typical values of $V_0$ can be even larger. Estimation $V_0>m_\pi/(2c_1)$ with $V_0\sim 0.2 m_\pi$ yields $c_1> 2.5$ and $\nu\gsim 20$.
At these conditions even a not too rapid rotation may result in the formation of a  condensate pion-sigma field in the form of a supervortex or a more whimsical vortex structure, provided  the vortex field is not destroyed by the temperature effects, which are disregarded in our present study.

(d) {\em Formation of vortices in magnetic field.}
In Ref. \cite{Zahed} the rotation of the vacuum of non-interacting charged pions was considered in presence of a rather strong  external uniform constant magnetic field $H$.  This analysis  becomes not applicable for $|e|H\lsim 1/R^2$, i.e. when the Larmor radius of the particle becomes larger than  the size of the vessel. The number of permitted states in the uniform magnetic field is given by $N=|e|HS/(2\pi)=|e|HR^2/2$ and for $|e|H< 2/R^2$ we have $N<1$.  Thus the results \cite{Zahed} do not describe  the  case $H\to 0$, which we have studied above.

 In our case the interaction with the magnetic field can be introduced within the model 1. Strong magnetic fields with $H\lsim m^2_\pi$,  $m^2_\pi/|e|\simeq 3.5\cdot 10^{18}$G, can be generated in  peripheral heavy-ion collisions and central regions of neutron stars \cite{Voskresensky:1980nk}. Uniform magnetic field inside the rotating ideal vessel can be generated, if the vessel is put inside a solenoid.  A magnetic field  can be generated also, if we deal with  the charged rotating cylindrical capacitor. In the latter case a simple estimate shows that for  $R=1$ cm it is  sufficient to switch on a tiny  external field $|e|H>10^{-8}$G
in order to get $N>1$ and thereby to overcome the problem with absence of the  solution $\mu_\pi =0$ of
Eq. (\ref{critOmvac}) at $V_0=0$. Now, in presence of such a  magnetic field satisfying condition $N>1$ instead  of Eq. (\ref{grlev}) we have
\be
\epsilon_{1,\nu}=-\Omega\nu -V_0+g\omega_0+\sqrt{\widetilde{m}^2+|e|H}\label{Heps}
\ee
and instead of Eq. (\ref{critOmvac}) we obtain
\begin{eqnarray} &\nu \Omega_c^H=\nu \Omega (\epsilon_{1,\nu}=0) \nonumber\\&=-V_0+g\omega_0+\sqrt{\widetilde{m}^2 +|e|H}\simeq -V_0+g\omega_0+\widetilde{m} \,.\label{critOmvacH}\end{eqnarray}
The latter relation holds  for $|e|H\ll m_\pi^2$. Please  compare (\ref{critOmvacH}) and (\ref{Omcrsmall}). We see that  dependence on $R$  disappeared from this relation, as in case studied in Ref. \cite{Zahed}. The degeneracy factor $0<\nu\leq N$.
The fields $H\lsim (10^5-10^7)$G  can be successfully generated at the terrestrial laboratory conditions. Note that   for $|e|H\sim 10^6$ G at $R=1$ cm we  have $N\sim %Rm_\pi\sim
10^{14}$. With $\nu=c_1 \widetilde{m}R$   we estimate that $\nu\lsim N$  for $c_1\lsim 10$ and $\Omega_c^H\lsim 10^{9}$ Hz.

(e) {\em Injection of the proton gas in rotating vessel.}
 In absence of the capacitor, in case of the rotation of the electrically neutral   vessel, in which  an amount of heavy positively charged particles is injected (e.g. protons, which are  7 times heavier than pions), the positive  charge density $n_p$ can be compensated by the produced  negatively charged pion vortex field, i.e. $|\rho_\pi|=\bar{\mu}\Phi^2_\pi \simeq n_p$.
 For $\mu_\pi =0$ the pion field energy  is ${\cal{E}}_{\pi}\simeq V_0 Z$, as it follows from the first line of Eq. (\ref{Eg0}). Maximum value of  $\Phi^2_\pi$   at $eHR^2\sim Ze^2\Omega R\gg1$ corresponds to $\bar{\mu}\simeq \widetilde{m}$, cf. ({\ref{critOmvacH}), and thus
 the minimum of the energy is given by ${\cal{E}}_{\pi}=(\widetilde{m}-\Omega\nu+g\omega_0) Z$ and   it becomes negative for $\Omega>(\widetilde{m}+g\omega_0)/\nu$, where the latter quantity coincides with $\Omega_c (V_0=0)$  given by Eq. (\ref{Omcrsmall}).

 Concluding, above we demonstrated that in the model 1 for $V=H=g\omega_0=0$ and in the model 2 for $g\omega_0=0$ the vacuum in the  rotating frame remains to be stable respectively  production of non-interacting  charged and neutral pions. However within the model 1
 instability for production of non-interacting charged pions in the rapidly rotating frame occurs  already in presence of a weak external electric field,  cf. (\ref{condVom}), and/or magnetic field and charged defects. Also  instability   feasibly  appears already in absence of external electric field owing to spatial redistribution of the charge at formation of the vortex.

{\bf Instability of vacuum  in laboratory frame.}
Above we considered a piece of the rotating vacuum $r<R$ assuming that the rigidly rotating reference frame rotates with a fixed  angular velocity. Now let us focus attention on  the case (ii).  The angular momentum needed for formation of the vortex is taken from the bucket walls.
Employing Eq. (\ref{standEin-fin}) for the energy balance,
and  Eqs. (\ref{grlev}), (\ref{Eg0}) now at $\Omega =0$,  we recover the condition %(\ref{critOmvac})
for the appearance of the vortices in the laboratory frame:
 \begin{eqnarray}
&\delta{\cal{E}}=2\pi d_z \int_0^R rdr \rho_{\pi,\Phi}\epsilon_{1,\nu}[\nu,\Omega =0]-L_{\pi,\Phi}^{\rm lab}\Omega\nonumber\\
&=
 2\pi d_z \int_0^R rdr\rho_{\pi,\Phi}\nonumber\\&\times( -V_0 +g\omega_0 +\widetilde{m}\sqrt{1+j_{1,\nu}^2/(R^2 \widetilde{m}^2)}-\Omega\nu)<0,\end{eqnarray}
 which coincides with that we derived above considering the rotation frame, cf. Eq. (\ref{bosenrot}).

{\bf About dynamics of creation of the vortex field.}
It is important to notice that in case of the vacuum placed in a strong static electric field in absence of the rotation, the charged bosons
are  produced non locally  by the tunneling of particles from the lower continuum to the upper continuum. As we have mentioned, the typical time of such processes is exponentially large $\tau\sim e^{m_\pi^2/|eE|}/ m_\pi$ for $|eE|\ll m_\pi$. Another mechanism is a production of pairs locally near a wall placed in the vacuum (Casimir effect). The probability of such processes is still smaller than the mentioned probability of the tunneling.

In case of the rotation of the empty vessel the charged pion field can be produced in more rapid processes, in reactions with particles of the rotating wall of the vessel.
As one of possibilities to create  the vortex condensate, one may inject  inside the  vessel an admixture of protons, as we have mentioned. The protons  accelerated during the rotation of the system will then produce the radiation of the charged pion pairs, which further can form the vortex field.

\subsubsection{ Ideal pion gas with fixed particle number in rotating system}

 In case of the  ideal gas characterized by the dynamically fixed particle number  $N_\pi$, being put in a resting vessel,
the value $\Phi^2_{0\pi,\sigma}$  is found from the normalization condition $N_\pi={\mu}_{\pi,\Phi} \Phi^2_{0\pi,\sigma}[\Omega =0]\pi R^2d_z$ obtained by integration of Eq. (\ref{muPhi}). In the rotating frame
the constant $\Phi^2_{0\pi,\sigma}$  is found from the  condition
 \begin{eqnarray} &N_\pi = \Phi^2_{0\pi,\sigma}\int_0^R\widetilde{\mu} \chi^2 2\pi r dr d_z\nonumber\\&
 \simeq\bar{\mu}\Phi_{0\pi,\sigma}^2 \pi d_z R^2 J_{\nu+1}^2 (R/r_0^{\rm lin}[\bar{\mu}])\,,\label{tildemuN}\end{eqnarray}
 with $r_0^{\rm lin}$ and $\bar{\mu}$  given in  Eq. (\ref{rlin}). Equation (\ref{tildemuN}) yields the relation between the fixed (on a time scale under consideration) value $N_\pi$ and the constant value $\bar{\mu}$. The quantity  $\epsilon_{n,\nu}={\mu}_{\pi,\Phi}$ depends on $\Omega$ through the relation (\ref{rlin}).

The  vortex energy is given by Eq. (\ref{bosenrot}), where now $\Phi_{0\pi,\sigma}^2$ is determined by the condition of the fixed particle number (\ref{tildemuN}), i.e., ${\cal{E}}_{\pi,\Phi}[\Omega]=N_\pi \epsilon_{1,\nu}$. To understand will the gas be at rest or rotating with the angular velocity $\Omega$ we should compare ${\cal{E}}_{\pi,\Phi}[\Omega]$ and ${\cal{E}}_{\pi,\Phi}[\Omega=0]$. The minimal value of the  quantity ${\cal{E}}_{\pi,\Phi}[\Omega=0]$ corresponds to $\nu =0$ and for $\widetilde{m}R\gg 1$ is given by
\begin{eqnarray}{\cal{E}}_{\pi,\Phi}[\Omega=0]\simeq N_\pi[-V_0+g\omega_0+\widetilde{m}+j^2_{1,0}/(2R^2\widetilde{m})],\end{eqnarray}
% for $\lambda\to 0$. For a very low pion gas density, cf.
compare with Eqs. (\ref{Edynpiappr}), (\ref{genEPhi}). For $\nu\ll R\widetilde{m}$
we find that
\begin{eqnarray}
&{\cal{E}}_{\pi,\Phi}[\Omega]-{\cal{E}}_{\pi,\Phi}[\Omega=0]
\nonumber\\&\simeq N_\pi [-\Omega\nu +(j^2_{1,\nu}-j^2_{1,0})/(2R^2\widetilde{m})]<0\,,
\end{eqnarray}
for
 \be\Omega>\Omega_{c1}^{\rm id}(\nu)\simeq  (j^2_{1,\nu}-j^2_{1,0})/(2\nu R^2 \widetilde{m})
 %\nu\ln (R/r_0^{\rm lin})/(R^2 \bar{\mu}_{\pi,\Phi})
 \,.\label{Omcr1id}
\ee
The minimal value $\Omega_{c1}^{\rm id}(\nu)$ corresponds to  $|\nu|=1$.  Comparing (\ref{Omcr1id}) and (\ref{critOmvac})
%, (\ref{critvaluerot})
we see that $\Omega_{c1}^{\rm id}\ll \Omega_{c}(\lambda =0)$ at least for small values  $V_0-g\omega_0$, i.e. in the presence of a pion gas the vortices appear already at much smaller rotation frequencies than in case of the rotating vacuum. For the former case at $\widetilde{m}R\gg 1$ we deal with nonrelativistic rotation for $\Omega\sim \Omega_{c1}^{\rm id}(1)$. For a single vortex with $\nu=1$
we have
\be
\delta {\cal{E}}^{(1)} \simeq -  [\Omega-\Omega_{c1}^{\rm id}(1)]N_\pi\,.
\ee
In presence of $\nu$ single vortices, each  with $\nu=1$, the energy gain is $\delta {\cal{E}}=\nu \delta {\cal{E}}^{(1)}$.

In absence of the external rotation, as well as for $\Omega<\Omega_{c1}$,  production of vortices is energetically not profitable. However, if a vortex appeared  by some reason, it would survive   due to conservation of the winding number. In this case presence of a  vortex results in  a weak self-rotation of the Bose gas with the rotation velocity $\omega_{\rm self}\sim -\nu/(\widetilde{m}R^2)$. In our consideration performed above we assumed that $|\omega_{\rm self}|\ll \Omega_{c1}$.
Also, in case of the ideal gas under consideration, at the increasing rotation frequency the individual vortices may form the lattice. This possibility will be considered in the next section on example of the self-interacting fields.

\subsection{Self-interacting complex scalar fields in  rotating system}\label{sect-selfinter}

\subsubsection{Equation of motion, boundary conditions and energy}
We continue to use the symmetry breaking term in the form ${\cal{L}}_{\rm s.b.}^{(2)}$.
For $\lambda\neq 0$, $V_0,g\omega_0\simeq const$,   employing Eqs. (\ref{Phifieldform}), (\ref{piSigma1111}),
in the dimensionless variable $x=r/r_0^\lambda$, now with
\be
r_0^\lambda=(\lambda v^2_i +\bar{\mu}^2-\widetilde{m}^{2})^{-1/2}>0\,,\label{rlambda}
\ee
 where $v_i =0$ for the model 1 and $v_i =v^2$ for the model 2, we arrive at equation:
\be (\partial^2_x +x^{-1}\partial_x -\nu^2/x^2)\chi+\chi -\lambda\Phi_{0\pi,\sigma}^2(r_0^\lambda)^2\chi^3=0.\label{filvorteqdim11lambda}
\ee

In case of the model 1  we have $r_0^\lambda=r_0^{\rm lin}$.  For the pion gas at a low density and a small  $\Omega$ using (\ref{rlambda}) and (\ref{Edynpiappr}) we have $r_0^\lambda \simeq  \sqrt{m^*_{\pi}/(\rho_\pi\lambda})$.
%In case of vacuum, employing ${\mu}_{\pi,\Phi}$ from Eq. (\ref{grlev}) for $\mu_\pi\neq 0$  we obtain  $r_0^\lambda\simeq R/\nu\ll R$ for $\nu\gg 1$, whereas at $\mu_\pi =0$ we have $r_0^\lambda =1/\sqrt{\bar{\mu}^2-\widetilde{m}^2}$.
In case of the model 2, for $\lambda\gg 1$ we have
 $r_0^\lambda \simeq 1/\sqrt{\lambda v^2}$ both in case of the low density pion gas and for the vacuum.

Similarly to  Eq. (\ref{phisol}) we may introduce the quantity
\begin{eqnarray}&\Phi_{0\pi,\sigma} =\sqrt{[v^2_i + (\bar{\mu}^2-\widetilde{m}^{2})/\lambda]}\,\nonumber\\&\times\theta (v^2_i+(\bar{\mu}^2-\widetilde{m}^2)/\lambda),\label{Phich2}
\end{eqnarray}
which is the solution of Eq. (\ref{filvorteqdim11lambda}) at $x\to \infty$ corresponding to the boundary condition $\chi (x\to \infty)\to 1$.

If we are interested  in description of  the bulk region far away from the boundary, we can ignore the
influence of the boundary condition at $r=R$ for $R\gg r_0^\lambda$.
Thus we  may chose boundary conditions  $\chi (x\to 0)\to 0$ and $\chi (x\to \infty)\to 1$. Then the asymptotic solution of Eq. (\ref{filvorteqdim11lambda}) for $x\gg 1$ gives $\chi =1-\nu^2/(2 x^2)$, and for $x\to 0$ we get $\chi \propto x^{|\nu|}$. Thus  the field $\Phi_{\pi,\sigma}$ is expelled from the vortex core and the equilibrium value (\ref{Phich2})
 is recovered at $r\gg r_0^\lambda$. An interpolation solution satisfying both asymptotics can be presented for $\nu>2$ as
\be \chi =x^{|\nu|}/[1+x^{|\nu|}(1+\nu^2/(2x^2))]\,.\label{interpolationsol}\ee

 We still should clarify how  the boundary condition at $r=R$  can be fulfilled on the wall of the vessel. As example, let us consider the model 1. We can solve Eq. (\ref{filvorteqdim11lambda})   employing  the variable $y=(r-R)/r_0^\lambda$,  $x= y+R/r_0^\lambda$,  for $R\gg r_0^\lambda$ at the boundary conditions
$\chi (y\to -\infty)=1$ and $\chi (y=0)=0$. The latter condition demonstrates absence of the normal component of the  $\Phi_{\pi}$ field flux through the boundary $r=R$.
At $r_0^\lambda\ll R$ for typical dimensionless distances $y\sim 1$, the angular momentum  term, $\sim (\nu r_0^\lambda)^2 /R^2$, and  the curvature term, $\sim r_0^\lambda /R\ll 1$, can be dropped for not too large $\nu$, which  means that  geometry can be considered as effectively one-dimensional one, cf. a similar argumentation employed in \cite{Migdal:1977rn}.
Then appropriate solution gets  the form
\be \Phi_{\pi}=-\Phi_{0\pi} e^{i\nu\theta}\mbox{th}[(r-R)/(\sqrt{2} r_0^\lambda)], \,\, r_0^\lambda\ll r<R\,.\label{tangPhi}
\ee

 Using Eqs. (\ref{LphiOmpiV}), (\ref{LphiOm}) and Eq. (\ref{Eg0}), which we derived above for $\lambda =0$, now for  $\lambda\neq 0$ we find
\begin{eqnarray}
&{\cal{E}}_{\pi,\Phi}(\Omega) =\frac{\Phi_{0\pi,\sigma}^2d_z}{2} 2\pi\int_0^R r dr\chi\nonumber\\&\times\left[-\partial_r^2 -\frac{\partial_r}{r} +\frac{\nu^2}{r^2}-(\lambda v^2_i +\widetilde{\mu}^2-\widetilde{m}^{2})+\lambda\Phi^2_{0\pi,\sigma}\chi^2\right]\chi\nonumber\\
&-\lambda\frac{\pi\Phi_{0\pi,\sigma}^4d_z}{2} \int_0^R r dr\chi^4+\frac{\lambda v^4_i d_z\pi R^2}{4}\nonumber\\&-\int_0^R n_p V 2\pi r dr d_z+{\Phi_{0\pi,\sigma}^2d_z} 2\pi\int_0^R r dr\chi^2\mu_{\pi,\Phi}^2\nonumber\\
&+\mu_{\pi,\Phi} (\Omega\nu+V_0-g\omega_0){\Phi_{0\pi,\sigma}^2d_z} 2\pi\int_0^R r dr\chi^2\nonumber\\
&={\cal{E}}_{\pi,\Phi}^{\rm l}+\frac{\lambda d_z}{4} 2\pi\int_0^R r dr
(v^2_i-\Phi_{0\pi,\sigma}^2\chi^2)^2\label{Eg}\\
%&=-\lambda\frac{\Phi_{0\pi,\sigma}^4d_z}{4} 2\pi\int_0^R r dr\chi^4+\frac{\lambda v^4_i d_z}{4}\pi R^2\nonumber\\
&+\frac{1}{2}{\Phi_{0\pi,\sigma}^2d_z} 2\pi\int_0^R r dr\chi^2\mu_{\pi,\Phi}\widetilde{\mu}_{\pi,\Phi}-\int_0^R n_p V 2\pi r dr d_z\,.\nonumber
\end{eqnarray}
 For $\nu =0$, $\chi =1$, $V_0=g\omega_0=0$ from (\ref{Eg}) we recover Eqs. (\ref{Edynpi}) and (\ref{genEPhi}). For  $\lambda =0$ we recover Eq. (\ref{Eg0}). Actually  the linearized equation of motion is recovered at a weaker condition $\chi^2\ll 1$.

Employing (\ref{Eg}) and equation of motion (\ref{filvorteqdim11lambda})
we find
\begin{eqnarray} &{\cal{E}}_{\pi,\Phi}(\Omega) =2\pi d_z \int_0^R r dr\label{Egcor}\\
&\times\left[-\frac{\lambda (v^2_i-\Phi_{0\pi,\sigma}^2\chi^2)^2}{4}
-\frac{\lambda v_i^2(\Phi_{0\pi,\sigma}^2\chi^2-v^2_i)}{2}\right.\nonumber\\ &\left.+\mu_{\pi,\Phi}\widetilde{\mu} \Phi_{0\pi,\sigma}^2\chi^2 -n_p V\right]\,.\nonumber
\end{eqnarray}
For $V_0\simeq const$ and $g\omega_0=const$ that we assumed, we have
\begin{eqnarray}
&{\cal{E}}_{\pi,\Phi}\simeq 2\pi d_z \int_0^R r dr\nonumber\\&\left[-\frac{\lambda (v^2_i-\Phi_{0\pi,\sigma}^2\chi^2)^2}{4}
-\frac{\lambda v_i^2(\Phi_{0\pi,\sigma}^2\chi^2-v^2_i)}{2}+\widetilde{\mu}^2 \Phi_{0\pi,\sigma}^2\chi^2 \right]\nonumber\\
&-L^v\Omega -(V_0-g\omega_0) N_\pi +V_0 Z
\,.\nonumber\end{eqnarray}
Employing Eq. (\ref{Phich2}) we may also rewrite Eq. (\ref{Egcor}) as
\begin{eqnarray}
&{\cal{E}}_{\pi,\Phi}\simeq 2\pi d_z \int_0^R r dr\nonumber\\&\times\left[\frac{\lambda v_i^4(1-\chi^4)}{4}-\frac{v_i^2\chi^4 (\bar{\mu}^2-\widetilde{m}^2)}{2}+(\widetilde{\mu} -\Omega\nu - V_0+g\omega_0)\widetilde{\mu}v_i^2\chi^2\right]\nonumber\\
&+2\pi d_z \int_0^R r dr\label{Egcor1}\\&\times\left[-\frac{(\bar{\mu}^2-\widetilde{m}^2)^2\chi^4}{4\lambda}+\frac{(\widetilde{\mu} -\Omega\nu - V_0+g\omega_0)\widetilde{\mu}
(\bar{\mu}^2-\widetilde{m}^2)\chi^2}{\lambda}-n_p V\right].\nonumber\end{eqnarray}
In the model 1 there remains only  contribution of the second line. In the model 2 the latter contribution  can be dropped for $\lambda\gg 1$ compared to the contribution of the first line.

\subsubsection{Rotating vacuum. Empty rotating vessel}
{\bf Instability of vacuum in rotating frame, case (i).}
For $\mu_{\pi,\Phi} =\bar{\mu} -\Omega\nu - V_0+g\omega_0=0$ from (\ref{Egcor1})
%and (\ref{Phich2})
for $ v^2_i+(\bar{\mu}^2_{\pi,\Phi}-\widetilde{m}^2)/\lambda>0$ we have
\begin{eqnarray}
&{\cal{E}}_{\pi,\Phi}(\Omega)=2\pi d_z \int_0^R r dr\label{envplm}\\&\times\left[-n_p V+\frac{\lambda v_i^4(1-\chi^4)}{4}-\frac{v_i^2\chi^4(\bar{\mu}^2-\widetilde{m}^2)}{2}-
\frac{(\bar{\mu}^2-\widetilde{m}^2)^2\chi^4}{4\lambda}\right]
%\nonumber\\&\times \theta (\lambda v_i^2 +(\Omega\nu+V_0)^2-\widetilde{m}^2)
.\nonumber
\end{eqnarray}
For $\chi =1$ we obtain
\begin{eqnarray}
&{\cal{E}}_{\pi,\Phi}(\chi =1)= \pi R^2 d_z \label{chi1}
\\
&\times\left[n_p V_0-\frac{v_i^2 ((\Omega\nu+V_0-g\omega_0)^2-\widetilde{m}^2)}{2}-
\frac{((\Omega\nu+V_0-g\omega_0)^2-\widetilde{m}^2)^2}{4\lambda}\right].
%%\nonumber\\ &\times\theta (\lambda v_i^2 +(\Omega\nu+V_0)^2-\widetilde{m}^2)
%+{\cal{E}}_{\rm kin}^{(1)}
\nonumber\end{eqnarray}
In the model 1   only the last term in Eqs. (\ref{envplm}), (\ref{chi1})
remains for $n_p=0$. So  production of the vortex condensate becomes to be  energetically favorable, ${\cal{E}}_{\pi,\Phi}(\Omega)<0$, for $\Omega\nu+V_0-g\omega_0>\widetilde{m}$, i.e. for
\be\Omega>\Omega_c^{\pi} =(\widetilde{m}-V_0+g\omega_0)/\nu\,.\label{omcr1mod1}\ee
Note that  $\Omega_c^{\pi}$ approximately coincides with $\Omega_c$ given above by Eq. (\ref{Omcrsmall}) and (\ref{critOmvacH}) but differs from (\ref{critOmvac}).
 The quantity $\Omega_c^\pi$ can be made very small for very large values of the quantum number $\nu$ provided $\widetilde{m}-V_0+g\omega_0>0$, certainly at the condition that
 in case (ii) the maximum value of $\nu$ is restricted by the value of the initial rotation angular momentum.

 On the other hand it is important to notice that the asymptotic solution $\chi =1-\nu^2/(2x^2)$  is valid only for $R\gg \nu r_0^\lambda$, i.e. for $\Omega \nu +V_0-g\omega_0\gg \sqrt{ \widetilde{m}^2+ \nu^2/R^2}$.  Thus this asymptotic solution is  realized for $\Omega R<1$ only for $\Omega$ very near the limiting value $1/R$  for $V_0-g\omega_0 \gg \widetilde{m}^2 R/(2\nu) $ at  $\nu\gg \widetilde{m}R$, cf. Eq. (\ref{condVom}) for non self-interacting pions.

In the model 2 the second term in the square brackets in Eq. (\ref{envplm}) produces  the contribution to the kinetic energy  given by
\be{\cal{E}}_{\rm kin}=2\pi d_z
%\nu^2\Phi_{0\pi,\sigma}^2 d_z\pi\int_{r_0^\lambda}^Rdr\chi^2/r\simeq \nu^2\Phi_{0\pi,\sigma}^2 d_z\pi\ln (R/r_0^\lambda)
(\nu^2 v^2/2) \ln (R/\nu r_0^\lambda)) \label{ekin1}\,\ee
with a logarithmic accuracy. Here we assumed that asymptotic solution holds for $R/\nu r_0^\lambda\gg 1$  and $r_0^\lambda\simeq 1/\sqrt{\lambda v^2}$  for $\lambda\gg 1$, i.e. for $\nu\ll R\widetilde{m}\sqrt{\lambda}$.
Comparing the term (\ref{ekin1}) with the third term in square brackets in Eq. (\ref{envplm})
%, (\ref{chi1})
  we see that the  $\sigma\pi^0$ condensate vortex field can  appear for
\be \Omega >\Omega_c^{\Phi}=(\sqrt{\widetilde{m}^2+ 2\nu^2 \ln (R/\nu r_0^\lambda)/R^2}+g\omega_0)/\nu\,.\label{omegac22}\ee
As it is seen from this expression, there is no solution for $\Omega R<1$ for $g\omega_0=0$. Such a solution could exist only at not too small values of the attractive interaction $g\omega_0$.
Please compare Eqs. (\ref{omcr1mod1}) and  (\ref{omegac22}), which we derived here for the case of the  self-interacting fields, with Eq. (\ref{critOmvac})
%and (\ref{critvaluerot})
derived above for the case of not self-interacting fields.

For $\Omega >\Omega_{c}^{\pi,\Phi}$ the state $\sigma =v$, $\pi_i =0$  becomes  not a ground state provided conditions (\ref{omcr1mod1}) or (\ref{omegac22}) are fulfilled for $\Omega R<1$.  Near the walls of the rotating vessel  there  may appear numerous vortices and anti-vortices. Then vortices migrate into the vessel volume and antivortices are absorbed by the walls of the vessel. We recall that in case (i) the constancy of the rotation frequency is recovered from the external source of the rotation.

{\bf{Formation of vortices  in case (ii).}} In this case   a decrease of the angular momentum of the vessel can be energetically preferable. Therefore we should study this possibility similarly to that we have done in case of non self-interacting bosons.

Let us consider a nonrelativistic rotation. Employing condition (\ref{standEin-fin}) and Eq. (\ref{Lvortsingle1}) for $\mu_{\pi,\Phi}=0$  (in case of the vacuum) we obtain
\begin{eqnarray}
 &\delta {\cal{E}}\simeq  {\cal{E}}_{\pi,\Phi}^{\rm lab}-L_{\pi,\Phi}\Omega = {\cal{E}}_{\pi,\Phi} [\nu,\Omega =0] \nonumber\\
 &-\nu 2\pi d_z \int_0^R r dr (\Omega\nu+V_0-g\omega_0)\Phi^2_{0\pi,\Phi}\chi^2 \Omega  <0\,.\label{vortcondlambda}\end{eqnarray}
Within the model 1 the condition (\ref{omcr1mod1}) for appearance of the vortex field, i.e. $\Phi^2_{0\pi,\Phi}>0$ at $\lambda\gg 1$,  does not change. Within the model 2 presence of the additional term $-L_{\pi,\Phi}\Omega$ allows to overcome the causality problem only for sufficiently large negative values of $g\omega_0$, cf. Eq. (\ref{omegac22}).

The maximum value of the angular momentum of the vortex, $\nu_{max}$, is limited by the value of the angular momentum of the rotation of the vessel, $L_{in}$, given by Eq. (\ref{Lin}).

{\bf About dynamics of creation of the vortex field.}
In the ground state $\sigma =v$, $\pi_i =0$ of the static vacuum (at $\Omega =0$)  there is no any friction.   Charged $\pi^{\pm}$ pairs can be produced, e.g.,  in uniform static   electric field in the process of the tunneling of particles from the lower continuum to the upper continuum, however the probability of the production of pairs $e^{-{m}^2_\pi/|eE|}$ is tiny for the strength of the electric field $E\ll m^2_\pi/|e|\simeq 10^{21}$ V$/$cm, as we have discussed above, cf. \cite{Voskresensky2021QED}. In presence of a moving wall,  there arises  a tiny friction force between virtual particles  and the wall (dynamical Casimir effect, cf. \cite{Kardar1999}).  Also there are other reasons, which may cause  creation of vortices near the wall, e.g., interaction of the virtual charged pions with the electric charge of the particles of the wall. As in case $\lambda\to 0$, which we have discussed above, the most efficient way to produce a vortex field inside the rotating vessel might be is to put in it a rare rotating  gas of protons. The pions forming the vortex state  will be then produced in radiation reactions. Such processes do not require any tunneling of particles from the lower to upper continuum and their production is not suppressed by a $e^{-{m}^2_\pi/E}$ factor.

\subsubsection{Rotating supercharged nucleus}
Above we have reminded the idea of a possibility of existence of stable supercharged pion condensate nuclei. If we deal with a rotating  nucleus of a large atomic number $A\simeq 2Z$  for
$ n_p=\rho_0/2$,
there may appear the charged pion condensate compensating the initial proton charge in interior of the nucleus. For $Z|e^3|\gg 1$ the charge is repelled to  a narrow surface layer and  screened to a value $Z_s\sim Z/(Z|e^3|)^{1/3}$, cf. \cite{Migdal:1977rn}. In spite of this,  the condition $N=|e|H(0) R^2/2\gg 1$ with $eH(0) R^2 \sim Z_s e^2\Omega R$ is satisfied.

For $\lambda \to 0$, $g\omega_0=0$,
using (\ref{Heps}) for $\epsilon_{1,\nu}=0$ and  $|e|H\ll m^*_\pi$ and the second line (\ref{Eg0}) we have $V_0\simeq m^*_\pi -\Omega\nu$ and
\be {\cal{ E}}-{\cal{E}}_{in}\simeq (m^{*}_\pi -\Omega\nu  -32 \mbox{MeV})Z\label{freesuperchargedRot},\ee
cf. Eq. (\ref{freesupercharged}).
Vortex condensate state appears for ${\cal{ E}}-{\cal{E}}_{in}<0$, i.e., for $\nu/R>\Omega \nu =m^{*}_\pi  -32 \mbox{MeV}$.

In a realistic case,   $\lambda \simeq 20$,  instead of (\ref{selfsupercharged}) we now have
\begin{eqnarray} &{\cal{ E}}-{\cal{E}}_{in}\simeq  [\lambda\rho_\pi^2/(4m^2_\pi)  +m_\pi-\Omega\nu -32 \mbox{MeV}]Z\nonumber\\&\simeq  (m_\pi -\Omega\nu)Z\,.\label{selfsuperchargedRot}\end{eqnarray}
At a fixed $\Omega >m_\pi/\nu$  a rotating  nucleus forming a charged pion-sigma  supervortex is stable.

Let us recall estimates of Ref. \cite{Xu-Guang Huang} done for the values of the angular momentum, $\nu \lsim 10^6$, and rotation frequency, $\Omega \simeq 0.05 m_\pi$, performed for peripheral heavy-ion collisions at $\sqrt{s}=200$ GeV. Thus we may expect occurrence of  metastable rotating  states in peripheral heavy-ion collisions.
Also in case of   rotating  cold superheavy nuclei  and  nuclearites
%at the condition  $\Omega\nu>m_\pi$, i.e. $\Omega R>1/c_1$ at $\nu =c_1 m_\pi R$,
it might be profitable to form a charged pion   vortex condensate, which will stabilize them in the rotating frame. The kinetic energy of such a rotating nuclear systems is then lost on a long time scale via a surface electromagnetic radiation. For very large number of baryons, $A$, such a radiation is strongly suppressed.

\subsubsection{Non-ideal gas with fixed particle number in rotating system}\label{lattice}
Now let us consider the case $\mu_{\pi,\Phi}\neq 0$ and $\mu_{\pi,\Phi}\gg \Omega\nu+V_0-g\omega_0$. This case is similar to that takes place at a nonrelativistic rotation
of cold atomic gases and He-II when $\mu_{\pi,\Phi}\simeq m_{\rm He}$ and $\Omega\nu+V_0-g\omega_0\ll m_{\rm He}$.

In absence of the  rotation of the vessel  appearance of vortices is  energetically not profitable, since the kinetic energy of the vortex with $\nu\neq 0$ is positive.   Moreover, at $\Omega =0$ the vortices characterized by $\nu >0$ could  be produced only in pairs with anti-vortices characterized by $-\nu$ due to the  angular momentum  conservation.   If a vortex having the integer winding number   $\nu$ was formed by some reason, it   would continue to exist till a collision with the corresponding antivortex, or with the walls of the vessel due to conservation of the angular momentum.

In presence of the rotation, in the rotation frame, using Eq. (\ref{Egcor}) and the asymptotic solution $\chi =1-\nu^2/(2 x^2)$ of the equation of motion for $x\gg \nu$ we find that the energy balance  is controlled by the kinetic energy of the vortex, ${\cal{E}}_{\rm kin}^{(1)}$, given by the first term in the second line (\ref{Egcor1}) in the model 1 and by the first term in the first line (\ref{Egcor1}) in the model 2, and the rotation contribution $L\Omega$ extracted from the last term in squared brackets in first line (\ref{Egcor1}). The same consideration can be performed in the laboratory frame employing Eq. (\ref{standEin-fin}). In the latter case  the  kinetic energy associated with the single vortex line  with the logarithmic accuracy is given by
\be{\cal{E}}_{\rm kin}^{(1)}\simeq\frac{\int d^3 X |\nabla \Phi_{\pi,\sigma}|^2}{2} =\frac{d_z \pi \nu^2 \rho_{\pi,\Phi}}{\bar{\mu}}\ln (\widetilde{R}/{r_0}^\lambda)
%\left(v^2+\frac{(\mu +\Omega \nu)^2-m^{*\,2}}{\lambda}\right)
%%\simeq d_z \pi \nu^2 (v^2+O(1/\lambda))\ln (\widetilde{R}/{r_0})
.\label{Evortex1}\ee
At large distances $r$ we cut   integration at $r\sim \widetilde{R}\gg r_0^\lambda$, being   the transversal size of the vessel $R$ in case of the single vortex line with the   center at $r=0$, and at the distance $R_L$ between vortices in case of the lattice of vortices. The latter possibility will be considered below. At small distances  integration is naturally cut at $r\sim r_0^\lambda$.

{\bf Lower critical angular velocity.}
Let us consider  the system at approximately constant density $\rho_{\pi,\Phi}$.  Then from the condition ${\cal{E}}_{\rm kin}^{(1)}-\vec{L}\vec{\Omega} <0$  we now find that the first vortex filament (together with the antivortex) appears for
\be
\Omega >\Omega_{c1}^\lambda (\nu)=\frac{\nu\ln ({R}/{r_0^\lambda})}{R^2 \bar{\mu}}\simeq \frac{\nu\ln ({R}/{r_0^\lambda})}{R^2 \widetilde{m}}\,.\label{Omcr1}
\ee
In the last equality we used that for a low density and for a slowly rotating gas $\bar{\mu}\simeq \mu_{\pi,\Phi}\simeq \widetilde{m}$.
Please compare this result with Eq. (\ref{Omcr1id}) derived above for the case of the ideal gas.
We should put $\nu =1$ and take $\widetilde{R}$ to be equal to the  maximum distance between the vortex and the edge of the vessel  $\sim R$, since it corresponds to the minimum value of $\Omega_{c1}^\lambda= \Omega_{c1}^\lambda(\nu =1)$.

{\bf Landau critical velocity for formation of vortices.} For a   vessel of a large size, $R\gg r_0^\lambda$,  following (\ref{Omcr1}) we have  $\Omega_{c1}^\lambda R\ll 1$.
The quantity $u_{v,\rm L}=\Omega_{c1}^\lambda R<1$ can be treated as the Landau critical velocity for formation of vortices, being very massive excitations compared to other particle  excitations. Due to this circumstance for  rotating systems of a large size $u_{v,\rm L}$ is much less than $u_{\rm L}$ necessary for a production of roton-like excitations with the momentum $k\neq 0$ occurring  in some rectilinearly moving and  rotating systems studied in
\cite{Pitaev84,V93jetp,Vexp95,BaymPethick2012,Kolomeitsev:2016isb,V93jetp,Kolomeitsev:2016isb}.

{\bf Interaction between vortex lines.} With the help of Eq. (\ref{Evortex1}) one may also recover the energy of the interaction between two vortex lines, cf. \cite{PethickSmith},
\begin{eqnarray}&{\cal{E}}_{\rm int}^{(2)}=\int d^3 X |\nabla \Phi_1\nabla\Phi^*_2+\nabla \Phi_1^*\nabla\Phi_2)/2\nonumber\\
, &\simeq d_z \pi \nu_1\nu_2 (\rho_{\pi,\Phi}/\widetilde{\mu})\ln (\widetilde{R}/{r_{12}})\,,\end{eqnarray}
 where $r_{12}$ is the distance between two vortices under consideration having momenta $\nu_1$ and $\nu_2$, $\widetilde{R}$ is here the distance from the vortex to the edge of the vessel at  $\widetilde{R}\gg r_{12}\gg r_0^\lambda$. This interaction energy is smaller than  the energy of two isolated vortex filaments, $2{\cal{E}}_{\rm kin}^{(1)}$ for $\nu_1=\nu_2=\nu$.

{\bf Spirals.}
A slightly deformed vortex line undergos a precession. Undergoing a long-wave oscillation the vortex line forms a spiral, cf. \cite{PitString}. Permitting a shift of the line in perpendicular direction  $x=d_{\perp}\cos (kz-\omega t)$, $y=d_{\perp}\sin (kz-\omega t)$ we have
\begin{eqnarray}
&\delta d_z =\int_0^{d_z}\sqrt{1+(\partial x/\partial z)^2+(\partial y/\partial z)^2}dz-d_z \nonumber\\&\simeq d_z d_{\perp}^2 k^2/2\,.\end{eqnarray}
Using Eq. (\ref{Evortex1}) we have
\be\delta {\cal{E}}_{\rm kin}^{(1)}\simeq d_z d_{\perp}^2 k^2 \pi \nu^2 \rho_{\pi,\Phi}/(2\bar{\mu})\ln (1/k{r_0}^\lambda)
\,,\label{Evortex11}\ee
where we assumed that $1/k\gg {r_0}^\lambda$, and
\be\delta {{L}}_v
%=\rho_{\pi,\Phi}\int_0^{d_{\perp}}r dr \oint \vec{v}d\vec{l}
\simeq -d_z d_{\perp}^2 \pi \nu \rho_{\pi,\Phi}\,.
\ee
The precession frequency $\omega_{\rm prec}$ is given by
\begin{eqnarray}&\omega_{\rm prec}=(\partial \delta {\cal{E}}_{\rm kin}^{(1)}/\partial d_{\perp})/(\partial \delta {\cal{L}}_z^{(1)}/\partial d_{\perp})\nonumber\\&=-\nu k^2\ln (1/k{r_0}^\lambda)/(2\bar{\mu})\,,\end{eqnarray}
i.e., the spiral rotates in opposite direction to the direction of the external angular velocity.

{\bf Rings.} In the system of a finite size  individual vortices characterized by the minimal angular momentum $\nu =1$ may form rings. Their energy is also given by Eq. (\ref{Evortex1}), however $d_z$ should be replaced by $2\pi R_{\rm ring}$, where $R_{\rm ring}$ is the radius of the ring provided $R_{\rm ring}\gg r_0^\lambda$. As it follows from Eq. (\ref{pphi}), the full momentum of the  ring is
 $p^{\rm ring}_\theta =\int_{r_0^\lambda}^{R_{\rm ring}} [\nu \rho_{\pi,\Phi} /r]2\pi r dr 2\pi R_{\rm ring}$ with a logarithmic accuracy, cf. \cite{PitString},  and thereby
 $$\vec{p}_{\rm ring}=4\pi^2 R_{\rm ring}^2 \rho_{\pi,\Phi}\nu \vec{e}_\perp\,,$$
 $\vec{e}_\perp$ is the unit vector perpendicular to the ring. In   the rotating frame  the vortex rings move with the momentum
 \begin{eqnarray} &P_{\rm ring}=\widetilde{\mu} \frac{d{\cal{E}}^{(1)}_{\rm kin}}{d{p}_{\rm ring}} = \widetilde{\mu}\frac{d{\cal{E}}^{(1)}_{\rm kin}/dR_{\rm ring}}{d{p}_{\rm ring}/dR_{\rm ring}}\nonumber\\&=\nu \frac{\ln (R_{\rm ring}/r_0^\lambda)}{4R_{\rm ring}}\,.\label{vring}\end{eqnarray}
 It is curious to notice that  giant  classical vortex rings with $\nu\gg 1$ can be formed in heavy-ion collisions and in rotating nuclei, cf. \cite{Becattini2013,Teryaev2017,Ivanov2023,Nesterenko2019} and references therein.

 {\bf Supervortex and vortex lines with $\nu =1$.} Minimization of the quantity
\begin{eqnarray} &\delta {\cal{E}} (\nu,\Omega)={\cal{E}}_{\rm kin}^{(1)}({R},\nu) -L^v({R},\nu)\Omega \nonumber\\&=-L^v({R},\nu) [\Omega -\Omega_{c1}^\lambda(\nu)]\label{Evnu1}
\end{eqnarray}
 in $\nu$ yields for ${R}\gg r_0^\lambda$:
\be\nu_m ={R}^2 \widetilde{\mu}\Omega/[2\ln ({R}/r_0^\lambda)]\,.\ee
Here we used Eqs. (\ref{Lvortsingle1}), (\ref{densmod1rot}) and (\ref{Evortex1}).
As we see from Eq. (\ref{Evnu1}), $d\delta{\cal{E}}/d\Omega\neq 0$ for $\Omega \to \Omega_{c1}^\lambda$, being in favor of the first-order phase transition at $
 \Omega = \Omega_{c1}^\lambda$.
In accordance with Eq. (\ref{Omcr1})  for $\Omega > 2\Omega_{c1}(\nu)$ we have $\nu_m>1$.
However comparison of $ \delta{\cal{E}}(\nu_m,\Omega)$ and $\nu_m {\cal{E}}(\nu =1,\Omega)$ shows that at  $\Omega>2\Omega_{c1}(\nu)$ the supervortex state becomes to be unstable respectively the decay on $\nu_m$ vortices with $\nu =1$ at least  for ${R}\gg r_0^\lambda$.

At a slow rotation, for an individual vortex  the circulation is $\kappa =\int \vec{v}d\vec{l}=2\pi|\nu|/\widetilde{m}$,  cf.  \cite{Tilly-Tilly}.
In a general relativistic case  with the help of Eq. (\ref{j}) we may write
\begin{eqnarray} &\int\vec{j}_{\pi,\Phi} d\vec{l}=\int (\Phi_{\pi,\sigma}^*\nabla \Phi_{\pi,\sigma} - \Phi_{\pi,\sigma}\nabla \Phi_{\pi,\sigma}^*)d\vec{l}/(2i)\nonumber\\&=\nu \rho_{\pi,\Phi} 2\pi/\bar{\mu}\,,
\end{eqnarray}
thus we recover relativistic generalization of the expression for the  circulation, i.e., $\kappa =2\pi \nu/\bar{\mu}$.

{\bf Lattice of vortices.} As we have mentioned, the vortices may form a   lattice and the system
begins to mimic   rotation of  the rigid body characterized by the linear velocity $v_{\rm rig}=\Omega R<1$.
In  case of a vortex lattice we have \cite{Tilly-Tilly}:
\be N_v^{\rm rig} \kappa = n_v \pi R^2 \kappa =2\pi R\cdot \Omega R\,.\label{Nvvort}\ee
Here   $N_v^{\rm rig}$ is the  total number of vortices inside the vessel of the internal radius $R$, which should be formed at given $\Omega$ in order the interior of the vessel would rotate as a rigid body together with the walls, and $n_v$ is the corresponding number of vortices per unit area.
We have
\be N_v^{\rm rig}=R^2/R_L^2\,,\quad n_v =1/(\pi R_L^2)\,,\label{Nn}\ee
 and thereby distance between vortices at a rigid-body  rotation,
\be
R_L=\sqrt{\nu}/\sqrt{\bar{\mu}\Omega}\,, \label{Rlat}
\ee
decreases with increasing $\Omega$.

The  energy gain due to the rigid-body rotation of  the  lattice of vortices mimicking the rotation of the vessel is given by  \cite{Tilly-Tilly,Hall},
\be
%\delta {\cal{E}}_{v}^\Phi
\delta {\cal{E}}\simeq N_v^{\rm rig} [{\cal{E}}_{\rm kin}^{(1)}(R_L) -L^v(R_L,\nu)\Omega]
%\simeq -N_vL(R_L)\Omega = \sqrt{\lambda/2} v^3 \Omega \pi R^2 d_z
%-{\cal{E}}_{in}
%c_1 MR^2\Omega^2
\,.\label{chirvortengain}\ee
This result  is obtained within a simplifying assumption of a uniform distribution of vortices \cite{BaymChandler1983}.  A more accurate result   computed for the triangular lattice \cite{Tkachenko,FischerBaym2003}, differs only by a factor $\frac{\pi}{2\sqrt{3}}\simeq 0.91$  from  that found for the uniform approximation. Also, following simplifying consideration of Ref. \cite{Tilly-Tilly} we disregarded a  difference of  the rotation angular velocity of the vortex lattice $\omega$ from that of the vessel $\Omega$. As it is shown in Appendix 2, for $R\widetilde{m}\gg 1$ this difference proves to be a tiny quantity.

%%%where ${\cal{E}}_{in}$ satisfies Eq. (\ref{EpsilonM})
%$c_1 =1/2$ in case of empty vessel of mass $M$ and $c_1 =1/4$ in case of the vessel uniformly filled by the matter and we used Eq. (\ref{Nvvort}) and put $\nu =1$.
Minimization of  (\ref{chirvortengain})  yields
\be N_v^{\rm rig} L_v(R_L,\nu)=N_v^{\rm rig}\pi\rho_{\pi,\Phi} \nu^2 d_z/(2\bar{\mu}\Omega)
%\pi d_z v^2 N_v^2\nu^2/(2\Omega)
\,.\label{Lvortlat}\ee
 %%%Similar expression follows from Eqs. (\ref{Lvortsingle}) and (\ref{Rlat}).
 Setting this expression to (\ref{chirvortengain}) and using (\ref{Nn}) and (\ref{Rlat}) we obtain
 expression for the equilibrium  energy
 \be \delta {\cal{E}}
 %{\cal{E}}_{v}^\Phi
 \simeq \rho_{\pi,\Phi}  \nu \Omega \pi R^2 d_z[\ln (R_L/r_0^\lambda) -1/2]
 %-{\cal{E}}_{in}
 %-c_1 MR^2\Omega^2
 \,.\label{latdeltae}
 \ee
Minimum of $\delta {\cal{E}}$ corresponds to $\nu =1$.

Thus, on this   example we demonstrated that at the rotation frequency $\Omega >\Omega_{c1}^\lambda$  in the rotating vessel filled by a pion-sigma gas (in our case at $T=0$) in cases described by both models 1 and 2 there may appear chiral-vortices, which at increasing $\Omega$ may form  the lattice mimicking the rigid-body rotation.

{\bf About upper critical angular velocity.} With a further increase of the  rotation frequency
the lattice is destroyed. The  minimal distance  $R_L \sim r_0^\lambda$ at a dense packing of vortices  corresponds to the number of vortices per unit area $n_v\sim 1/(\pi (r_0^\lambda)^2)$ in Eq. (\ref{Nn}) and to the maximum rotation frequency  given by
\be\Omega \simeq \Omega_{c2}\sim 1/[(r_0^\lambda)^2 \widetilde{m}]\,.\label{Omc2}\ee
For a larger rotation frequency, $\Omega >\Omega_{c2}$, the $\Phi_{\pi,\sigma}$ vortex-state should disappear completely and the initial $\sigma =v$, $\pi_i =0$ vacuum state is restored.
Note that for an extended rotating system the value $\Omega_{c2}R\gg 1$, however now it does not contradict to causality, since the system does not anymore rotate as a rigid body but it is separated on filaments with typical distance $\sim r_0^\lambda$ between them and $\Omega_{c2}r_0^\lambda\ll 1$. Note that in cold atomic gases the breakup of the lattice occurs when $\Omega$ reaches the value  $\Omega_h\ll \Omega_{c2}$,  cf.  Ref. \cite{FischerBaym2003}, $\Omega_h \sim \Omega_{c2} \xi/R$ in our case $\xi =r_0^\lambda$,  and for $\Omega>\Omega_h$ in the center of the bucket there appears a hole.

\subsection{Some estimates}

Taking  $R=1$ cm we have $\Omega R<1$ for $\Omega <3\cdot 10^{10}$Hz.  In case of He-II,  $\Omega_{c1}\sim 0.01$Hz for $R=1$cm,  $\Omega_{c2}\sim 10^{12}$Hz. In $^{87}$Rb, $\Omega_{c2}\sim 10^{4}$ Hz.
For  $\mu_{\pi,\Phi} =m_\pi$ following Eq. (\ref{Omcr1}) and using that   $m_\pi \simeq 2.1\cdot 10^{23}$Hz we estimate $\Omega_{c1}^\lambda\sim 0.1$ Hz  at  $|\nu|=1$. For $\Omega \gg\Omega_{c1}^\lambda$ we have $L^v\Omega \gg {\cal{E}}_{\rm kin}^{(1)}$.

With the help of Eq.  (\ref{Nn}), (\ref{Rlat}) we may estimate number of vortices in the lattice at the rigid-body rotation
   \be
   N_v =\bar{\mu} \Omega R^2/\nu <\bar{\mu} (R/R_{>})^2 R_{>}/\nu\,,\label{Nnv}
   \ee
   where we used that $\Omega R_{>}<1$.
For $R_{>}=10$m, $R=1$cm, and $\bar{\mu} \sim m_\pi$, $\rho_{\pi,\Phi} \sim m^3_\pi$ we estimate $N_v<10^{10}/\nu$. The  distance between vortices is $R_L\gsim \sqrt{\nu/\Omega[\rm Hz]} 10^{-2}$cm. Causality condition for the rigid-body rotation is fulfilled for  $\Omega <1/R_{>}\sim 10^7$Hz.

The initial rotation energy  should be larger than the rotation energy of the condensate, i.e. the condition
  \be \nu\rho_{\pi,\Phi} \Omega R^2 < \rho_M \Omega^2 R^4_{>}/4\,\label{MPhi}\ee
 should be fulfilled in case if $R_{>}-R\gg R$.
From here we get $\nu\rho_{\pi,\Phi} \ll \rho_M (R_{>}/R)^2 R_{>}$, provided $\Omega R_{>}< 1$. Taking $\rho_M\sim 10$ g$/$cm$^3$ and $R_{>}=10$ m, $R=1$ cm we obtain $\nu\rho_{\pi,\Phi} \ll 10^9\rho_0$. So, the condition (\ref{MPhi}) is easily satisfied, since in any case in hadronic systems we deal with  $\rho_{\pi,\Phi} \lsim 10\rho_0$. Taking $R_{>}\sim R=1$cm we get $\nu\rho_{\pi,\Phi} \ll \rho_0$.
For $\rho_{\pi,\Phi}\sim 0.1 \rho_0$ with $\rho_M\sim 10$ g$/$cm$^3$ and $R_{>}=10$ m we have $\nu <10^{10}$ and we estimate  $\nu <10$ for  $R_{>}\sim R=1$cm.

Nucleons in pulsars form a superfluid. Then  we deal with the neutron Cooper pairs, which  play a role of the boson excitations,  and  $\bar{\mu}\simeq 2m_N$.
Taking $\nu =1$ and using Eqs. (\ref{Nvvort}), (\ref{Nnv}) one gets estimation $n_v\simeq 6.3\cdot 10^3 (P/\rm{sec.})^{-1}$ vortices$/$cm$^2$ provided rotation period $P$ is measured in seconds, cf.  Ref. \cite{Sauls1989}. Then for the Vela pulsar having period $P\simeq 0.083$ sec the distance between vortices is $~4\cdot 10^{-3}$ cm.
For the pion superfluid (at dynamically fixed particle number), $\bar{\mu}\simeq m_\pi$  and we get $n_v\simeq 5\cdot 10^2 (P/\rm{sec.})^{-1}$ vortices$/$cm$^2$.

As we have mentioned, rotation with a large circular frequency (certainly at the constraint $\Omega R<1$) is  possible in energetic peripheral heavy ion collisions. Taking diameter of the overlapping region of colliding nuclei to be $R = 10$ fm we get $\Omega <\Omega_{\rm caus}=1/R\simeq 0.14 m_\pi$. Employing Eq. (\ref{Omcr1}) we estimate $\Omega_{c1}^\lambda\sim 0.05 m_\pi$.
Taking   $\Omega \sim (10^{21}-10^{22})$ Hz, $R\simeq 10$ fm we estimate $N_v\sim 3-30$.
So, in a heavy-ion collision a part of the initial angular momentum could be transferred from the baryon subsystem to the  chiral-vortex structure.

 {{We should  notice that, as it is believed,  the spin polarization of  particles emitted in heavy-ion collisions is  induced by the coupling of the  angular momentum produced by colliding nuclei at a non-vanishing impact parameter and the spin of particles distributed in the matter, cf. the Barnett effect.
 Baryons in heavy-ion collisions participate  in production of strange particles,  e.g., $\Lambda$ hyperons.  The polarization of the $\Lambda$ hyperon is measured in its rest frame, cf. \cite{ITS,VKT} and references therein. In our case,  the pion vortices, which can be formed provided the temperature is smaller than the critical temperature for their production and $\Omega >\Omega_{c1}^\lambda$,  absorb a part of the angular momentum of the system and even may  mimic  a rigid-body rotation of the system.  At the freeze out they return part of their angular momentum back to baryons contributing to the baryon polarization and  thereby the measurable $\Lambda$ polarization. This possibility was not yet studied. Also, pion and baryon momentum distributions for particles  involved in  the vortex structures participating in the rotation  should be different from the ordinary thermal  distributions. }}

 {{ For completeness we should also notice that the particle vortex rings are inherent in the  fluid dynamics. An example of such vortex rings is the smoke rings.  The toroidal baryon vortex structures can be formed not only in heavy-ion collisions but also in proton-nucleus collisions, cf. \cite{Lisa}. Vortex rings produced by jets propagating through the quark-gluon matter were considered in Ref. \cite{Serenone}. In heavy-ion collisions two baryon vortex rings can be formed at the periphery of the stronger stopped matter in the central region, one at forward rapidities and another at backward rapidities, cf. \cite{Ivanov2023} and references therein.   At forward/backward rapidities signal is expected to be strongest. The ring $\Lambda$ polarization observable is $R_\Lambda =\langle \vec{P}_\Lambda \vec{\eta}\rangle_y$, where $\vec{P}_\Lambda$ is the polarization of $\Lambda$, $\vec{\eta}=[\vec{e}_z\times \vec{p}]/|[\vec{e}_z\times \vec{p}]|$, $\vec{e}_z$ is the unit vector along the beam, $\vec{p}$\, is the momentum, averaging  runs over all
momenta with fixed rapidity $y$.
Reference \cite{Ivanov2023} argued that the  vortex rings  can be detected by measuring the ring  observable $R_\Lambda$ even in rapidity range $0 < y < 0.5$ (or $-0.5 < y < 0$) on the level of $(0.5-1.5)\%$ at $\sqrt{s_{NN}} = (5-20)$ GeV. It would be  interesting to study baryon-pion interaction within the vortex structures. These questions however require  a detailed investigation, that goes beyond the scope of this paper.}}

\subsection{Nuclear medium effects on chiral vortices}\label{polarization-sect}

In the nucleon matter the pion mass should be replaced by the
 effective pion gap,
\be
\widetilde{m}^{2}\to \widetilde{\omega}^2 (k_0)=m^2_\pi  -\mu^2_\pi +k_0^2 +\Pi (\mu_\pi, k_0)\,,\ee
in the above derived equations, cf. \cite{Migdal1978,MSTV90,Voskresensky:2022gts,Voskresensky:2022fzk}. Here  $\hat{k}\varphi =\nabla \varphi/i $, $k_0(\rho)$ corresponds to the minimum of the squared effective pion gap,
%$\widetilde{\omega}^2 (k_0)\leq m^2_\pi$,
$\Pi(\mu_\pi, k)$ is the pion polarization operator. For $\Omega =0$ the minimum of $\widetilde{\omega} (k)$ at $k_0\neq 0$ appears for $\rho >\rho_{c1}$,  $\rho_{c1}\simeq (0.5-0.8)n_0$, cf. \cite{Voskresensky:1993ud,Voskresensky:2022fzk}. Inequality  $\widetilde{\omega}^2 (k_0)>0$ holds for the nucleon density $\rho <\rho_{c\pi}$, where $\rho_{c\pi}\sim (1.5-3)m_\pi$ is the critical density for  the pion condensation. For a nuclear droplet  with $\rho$ nearby the critical point the value $\Omega_c$ may become very small, $\Omega_c\sim \widetilde{\omega} (k_0)$.

 For $\rho>\rho_{c\pi}$ one has  $\widetilde{\omega}^{2}(k_0(\rho))<0$. In absence of the rotation within the $\sigma$ model the p-wave pion condensation in the cold matter was considered in \cite{Baym1975}
and for nonzero temperature in \cite{VM1982}, cf. also \cite{MSTV90}.
The nuclear droplet with  $\rho >\rho_{c\pi}$  may undergo   self-rotation, some rough estimates for this case were done in \cite{V93jetp}.

We should also note that in some models, like the Manohar-Georgi  model, cf. \cite{Voskresensky:2022gts}, the value  $\widetilde{\omega}^2 (k_0=0)$ may become negative for $\rho>\rho_{c\pi}^s$. A rough estimate yields   $\rho_{c\pi}^s \sim 2\rho_0$ demonstrating a possibility of the s-wave pion condensation besides the p-wave one.

\section{Conclusion}\label{concluding-sect}
In this work within the linear $\sigma$ model we studied possibilities of the formation of the charged $\sigma\pi^{\pm}$ and neutral $\sigma\pi^0$  condensate complex vortex fields in various systems:   in the rotating cylindrical empty vessel, in the vessel filled by the pion-sigma gas at the temperature $T=0$  (Bose-Einstein condensate) with a dynamically  fixed  (on the time scale under consideration) particle number,  as well as in case of rotating nuclear systems. In the  case of the vessel filled by a pion-sigma Bose-Einstein condensate an analogy  is elaborated with  the Bose-Einstein condensates in cold gases and the condensed  $^4$He. Various applications of the  results were discussed.

In Section \ref{Linear-sect}  two models  describing the $\sigma\pi^\pm$ condensate (model 1) and the $\sigma\pi^0$  complex field condensate (model 2) were formulated.
Presence of complex fields is required  to consider then  the vortex field configurations in the rotating charged and neutral systems. {{In the paper body we employ the  symmetry breaking term in the Lagrangian density in the form explicitly  demonstrating  presence of  the nonzero pion mass,  ${\cal{L}}_{\rm s.b.}^{(2)}=-\frac{m^{*2}_\pi \vec{\pi}^2}{2}$ in Eq. (\ref{Lsb1}). Other choice, using ${\cal{L}}_{\rm s.b.}^{(1)}=\epsilon \sigma$, is considered in Appendix 1. In the latter case in the model 2 the neutral complex field  appears at much higher value of the chemical potential, $\mu_\Phi$, than it occurs for the model employing ${\cal{L}}_{\rm s.b.}^{(2)}$. In the model 1 both choices lead to the same results in our consideration. Therefore we further focused our consideration on the choice ${\cal{L}}_{\rm s.b.}^{(2)}$.}} We  demonstrated that in the rest frame the vacuum state proves to be stable respectively  creation of the pion condensate fields. In case of the pion-sigma  gas with a dynamically fixed particle number at $T=0$,   comparing the energies in the models 1 and 2 we found conditions, at which either the solution of the model 1 or of the model 2 is energetically favorable.
When the density of $\pi^+$ plus density of $\pi^-$ equals to the density of $\pi^0$, the charged field condensate proved to be energetically favorable compared to the  neutral $\sigma\pi^0$ condensate. However for the case when densities  of $\pi^+$, $\pi^-$ and $\pi^0$  are equal, the  solution within the model 2 proved to be energetically favorable. Moreover, the latter solution appeared to be energetically favorable in the chiral limit when the effective pion mass $m^*_\pi$ is   artificially put zero.   Also, within the model 1  we found the  response of the $\pi^\pm$ condensate on  the presence of  charged massive particles such as protons.

In Section \ref{chir} we studied chiral fields, which can be formed within mentioned two models in the rotating systems. First we noticed  that the rotation at a constant angular velocity is introduced in the Klein-Gordon equation  in the rotating reference frame similarly to the constant electric potential, cf. Eq. (\ref{LphiOmpiV}). In case of the nonrelativistic rotation the rotation term equivalently  can be introduced in the Schrodinger and Klein-Gordon equations with the help of the Galilean  shift of the spatial variables, cf. Eq. (\ref{rotSchrod}).  In case of the relativistic rotation we then discussed a subtle question about relations between  the rotating frame, rigid-body rotation  and the causality condition. Then we considered a possibility of the formation of the chiral-field vortex condensates in an empty rotating electrically neutral vessel. Conditions for the formation of vortices in the rotating frame and in the laboratory frame were discussed. Two cases were studied: when the vessel rotates at  the fixed rotation frequency, case (i), and at  the conserved initial angular momentum, case (ii).

In subsection \ref{sect-ideal} we considered the case of the non self-interacting fields ($\lambda =0$), which is not specific for the $\sigma$ model. For the charged pions studied  within the model 1 we included interaction with the static electric field $V=eA_0$ and in both models 1 and 2 for a generality we included a possibility of  additional coupling of the complex boson field with the 0-component of a static  attractive external vector  field $\omega_0$. Putting both $V$ and $\omega_0$ fields zero, we demonstrated that  the  energy levels $\epsilon_{n,\nu}$ in the rotating vessel  do not reach zero at increasing rotation frequency $\Omega$ for $\Omega R<1$. (Equation (\ref{grlev}) has no solution $\epsilon_{1,\nu}=0$ for $V=g\omega_0=H=0$ at $\Omega R<1$, where $R$ is the internal radius of the vessel). Thereby we concluded that  for $\lambda, V,g\omega_0,H$ equal zero the rotated $\sigma\pi$ vacuum remains to be  stable relatively formation of the vortex field.
However presence of even a rather small external attractive  $V$ or/and $\omega_0$ fields may allow for the  energy levels with  $\nu\neq 0$, where $\nu$ is the quantum of the angular momentum, to reach zero at $1/R>\Omega>\Omega_c$.
In case of  the charged pions described by the model 1 we demonstrated that
%%%%%the required value of the electric potential can be produced itself together with %%the vortex field, as the response on the rotation of the system at  %%$\Omega>\Omega_c$. So, a sometimes used statement that the cold vacuum cannot rotate %%was questioned. Besides this,
the vortex boson fields can be formed in the rotating vacuum
in presence of external electric and magnetic fields at  $\Omega>\Omega_c$.
Since the uniform rotation acts as an artificial electric field it may cause a separation of the electric charges and formation of a charged boson field vortex together with  an electric and magnetic  fields.
The  critical rotation frequency, $\Omega_c$, was  evaluated for the limiting cases of not too large values of the angular momentum, $\nu\ll {m}_\pi R$, cf. Eq. (\ref{Omcrsmall}), and for   the supervortex state with $\nu\gg {m}_\pi R$, cf.  Eq. (\ref{condVom}). We discussed various possibilities to observe  supervortex states. One of these possibilities is to place  the rotating vessel inside a charged capacitor in order to produce an additional electric field potential inside the vessel. Moreover we discussed influence of the switching on and off  an external magnetic field  and   injection  of external charged massive particles (e.g. protons) inside the vessel.

Then we focused attention on a possibility of the formation of the $\sigma\pi$ vortices in the rotated ideal pion-sigma gas at $T=0$ with a dynamically fixed particle number. Here,   the vortices appear for
 $\Omega>\Omega^{\rm id}_{c1}(\nu=1)\sim 1/(R^2m_\pi)$ within both models 1 and 2, even in case $V_0=g\omega_0\to 0$. The latter quantity $\Omega^{\rm id}_{c1}(\nu=1)$ proved to be very small for $m_\pi R\gg 1$, cf. Eq. (\ref{Omcr1id}).  Formation of vortex lines, spirals and rings was considered.  With increasing $\Omega$ the individual vortices  form the lattice mimicking the rigid-body rotation of the system.

In subsection \ref{sect-selfinter} we studied rotation of the $\sigma\pi$ vacuum and the pion-sigma Bose-Einstein condensate with a dynamically fixed particle number in presence of the self-interaction. It was shown that in the realistic case of a large value of the coupling constant,  $\lambda$, the interpolation solution, cf. Eq.  (\ref{interpolationsol}), of the equation of motion (\ref{filvorteqdim11lambda}) differs significantly from the Bessel function, which holds in the limit $\lambda\to 0$, see solution (\ref{Bessel})  of Eq. (\ref{filvorteqdim1}).
The former solution is constant in a broad region outside a narrow vortex core.
Also, the question about the boundary condition at $r=R$ was discussed.
Then we focused attention on the consideration of the rotation of the vacuum in absence and in presence of the fields $V$ and $g\omega_0$.  In case of the model 1 we demonstrated that the value of the critical rotation velocity, $\Omega_c^\pi$, cf. (\ref{omcr1mod1}), coincides with the value $\Omega_c=\Omega (\epsilon_{1,\nu}=0, c_1\ll 1)$ given by
Eq. (\ref{Omcrsmall}) in case of non-self-interacting fields for $\nu\ll m_\pi R$ and with the value $\Omega_c^H$ given by Eq. (\ref{critOmvacH}) in presence of a magnetic field $H$ (for $2/R^2<|e|H\ll m^2_\pi$ both expressions (\ref{Omcrsmall}) and (\ref{critOmvacH}) approximately coincide). Note that Eq. (\ref{omcr1mod1}) is easily fulfilled  for any attractive $V$ and $g\omega_0$ at large $\nu$. So, a supervortex state is formed for $\Omega>\Omega_c^\pi$.
The largest possible values of $\nu$ allowed by conservation of the angular momentum are energetically favorable.
In case of the model 2,  condition for the appearance of the vortex, cf.  (\ref{omegac22}), proved to be essentially different from that found for non-self-interacting fields. The vortex solution for $\Omega R<1$  exists here only at sufficiently large  values of the attractive interaction $g\omega_0$.

We demonstrated that   inside the rotating superheavy nuclei of a large atomic number  consisting of the nuclear matter at a normal nuclear density at the condition   $\Omega R>1/c_1$, at $\nu =c_1 m_\pi R$, $c_1\gg 1$, it may become energetically  profitable to form a charged pion   supervortex. Thereby such a rotating nucleus proves to be stable  in the rotating frame (in case (i)). In case (ii) the kinetic energy of the rotating nucleus is lost on a  long time scale via the surface electromagnetic radiation. For very large number of baryons, $A$, such a radiation is strongly suppressed. So, it is worthwhile to seek rather long-living relativistically rotating nuclei in heavy-ion collisions, superheavy nuclei and nuclearites in cosmic rays.

We also indicated that in difference with the tunneling mechanism for the creation of the particle pairs in static and slowly varying  electric fields, the boson vortex field can be produced right inside the rotating vessel.
The most efficient way to produce a vortex field  might be is to inject into the vessel a rare gas of massive charged particles, e.g., protons. Charged pions forming the vortex state  will be then created in radiation reactions inside the system.

Then we studied behavior of the rotating nonideal pion-sigma Bose-Einstein condensate  (gas with a dynamically fixed particle number) at zero temperature. In this case the chemical potential
$\mu_{\pi,\Phi}\neq 0$ is found from the condition of the conservation of the particle number and we considered the case when $\mu_{\pi,\Phi}\gg \Omega\nu+V_0-g\omega_0$. Here our consideration is similar to that takes place at nonrelativistic rotation of the He-II and the cold atomic gases. We note that expression for the critical angular velocity, cf. Eq. (\ref{Omcr1}), in this case with a logarithmic accuracy coincides with Eq. (\ref{Omcr1id}) derived above for  the ideal gas.

Vortex lines may form spirals. In the system of a finite size,  individual vortices characterized by the minimal angular momentum $\nu =1$ may form rings.
Also we considered a possibility that the external angular momentum is  accumulated in the lattice of vortices mimicking   the rigid-body rotation. With increasing rotation frequency
the lattice is destroyed. As the result, either
a giant vortex state can be formed for $1/R>\Omega>\Omega_h$, or the system will represent a dense packing of individual vortices placed at distances $\sim r_0^\lambda$ from each other   undergoing a very rapid rotation (with $1/r_0^\lambda >\Omega >1/R$). For $\Omega >\Omega_{c2}$ the vortex fields should disappear completely.

We discussed various possible applications of the results to such objects as  the rotating empty buckets, buckets filled by bosons at a fixed particle number, nuclear rotating metastable objects and fireballs formed in peripheral heavy-ion collisions.
In the latter case the vortices may appear owing to a friction and viscosity between fluxes of hadrons-participants and hadrons-spectators.
One of the restrictions for the rapid rotation of the fireball  formed in a heavy-ion collision follows from the causality limit, $\Omega <\Omega_{\rm caus}$. For $\Omega \geq  \Omega_{\rm caus}\sim 1/R$, where $R$ now is the transversal radius of the nuclear fireball, one may expect occurrence of an instability. This instability may result in the formation of separately rotating vortices or a  fission of the system on smaller pieces rotating with higher rotation frequencies.
These possibilities should be studied more carefully.

In the given paper we studied vortex solutions  for complex scalar fields in rotating systems. The case of the vector complex boson fields will be considered elsewhere. Instabilities in rotating hot gluon plasma were recently studied by lattice simulations, cf.  \cite{Braguta:2023kwl}.

{\bf{Acknowledgments.}} Fruitful discussions with   E. E. Kolomeitsev, Yu. B. Ivanov and O. V. Teryaev   are acknowledged.

{{{\section{Appendix 1. Symmetry breaking term taken in form ${\cal{L}}_{\rm s.b.}^{(1)}$}\label{breaking1}}}
In this Appendix we study the case when the symmetry breaking term in the Lagrangian density  is taken in the form  ${\cal{L}}_{\rm s.b.}^{(1)}$}. For a small $\epsilon >0$ energy minimum corresponds to $\sigma =v+O(\epsilon)$.

From (\ref{EnerPhipi}) we see that the energy density of the state $\sigma =v+O(\epsilon)$ is now smaller than that for $\sigma =-v+O(\epsilon)$ by $\delta E \simeq -2v\epsilon$.
Substituting the fields  $\sigma^{\prime}=\sigma -v$, $\vec{\pi}^{\,\prime}=\vec{\pi}$ in the expression ${\cal{L}}_{ \phi}+{\cal{L}}_{\rm s.b.}^{(1)}$ and varying it in fields we find  new quantities
\be
\sigma =v+{\epsilon}/{(2\lambda v^2)}\,,
%\quad m_N=g(v+\frac{\epsilon}{2\lambda v^2})\,,
\quad m^2_\sigma =2\lambda v^2 +3\epsilon /v\,.\label{massessymbr}
 \ee
Replacing these quantities back to the Lagrangian density we see that  there appears the pion mass term  $\delta {\cal{L}}=-\epsilon \vec{\pi}^2/(2v)$ and thereby we find
  \be
  m^{*2}_\pi =\epsilon/v\,.\label{epsilonpi}
  \ee

Within the model employing the charged pion fields (model 1) we now take $\sigma =v+\frac{\epsilon}{2\lambda v^2}$, $\Phi_\pi =\pi_1+i\pi_2$, $\pi_3=0$. Including interaction of $\Phi_\pi$ with static electric field $V$ and static external field $\omega_0$ after separation of terms linear in $\epsilon$ we arrive to the  Lagrangian density
\begin{eqnarray} &{\cal{L}}_{\pi}^{V}=
\frac{(\mu_\pi -V-g\omega_0)^2|\Phi_{0\pi}|^2}{2}-\frac{|\nabla \Phi_{0\pi}|^2}{2}-\frac{m^{*2}_\pi |\Phi_{0\pi}|^2}{2}\nonumber\\
&-\frac{\lambda |\Phi_{0\pi}|^4}{4}+\epsilon v+\frac{(\nabla V)^2}{8\pi e^2}+n_p V\,,\label{phiVApp}\end{eqnarray}
compare with  Eq. (\ref{phiV}) taken at $\Phi_\sigma =0$. Thus all  results we obtain in the paper body employing the model 1 with the symmetry breaking term ${\cal{L}}_{\rm s.b.}^{(2)}$ hold also with ${\cal{L}}_{\rm s.b.}^{(1)}$.

Now let us focus on the  model 2. Shift of the vacuum value $\sigma$ by the linear term $\propto \epsilon$, which we did in order to reproduce pion mass term, results in necessity of  a modification of the model 2, where $\sigma$ and $\pi_3$ meson fields should be unified in a complex field. So, we   take
\be\Phi_\sigma =\widetilde{\sigma} +i\pi_3\,,
\ee
$\pi_1=\pi_2=0$, $V=0$, with the field $\widetilde{\sigma}=\sigma -v-\frac{\epsilon}{2\lambda v^2}$ counted from the new vacuum value. Then in the linear approximation in small $\epsilon$  we arrive at
\begin{eqnarray} &{\cal{L}}_{\Phi}=
\frac{(\mu_\Phi -g\omega_0)^2|\Phi_{0\sigma}|^2}{2}-\frac{|\nabla \Phi_{0\sigma}|^2}{2}-\frac{m^{*2}_\pi \pi_3^2}{2}\nonumber\\
&-\frac{m^{2}_\sigma \widetilde{\sigma}^2}{2} -\frac{\lambda |\Phi_{0\sigma}|^4}{4}+\epsilon v\,,\label{phiSigmaApp}\end{eqnarray}
Averaging of the term ${\frac{m^{*2}_\pi \pi_3^2}{2}+\frac{m^{2}_\sigma \widetilde{\sigma}^2}{2}}$ yields $\frac{M^{2} \Phi_{0\sigma}^2}{2}$ with $M^2=(m^2_\sigma +m^{*2}_\pi)/2\gg m_\pi^{*2}$. Thereby the complex mean-field solution $\Phi_{0\sigma}\neq 0$ in the model using ${\cal{L}}_{\rm s.b.}^{(1)}$ appears at much higher value of $\mu_\Phi$ than it occurs for the model employing ${\cal{L}}_{\rm s.b.}^{(2)}$.

\section{Appendix 2. Angular velocities of  vortex lattice and  vessel}
The kinetic energy density of the individual vortex in a rather rare  boson sigma-pion  gas at its nonrelativistic rotation with the angular velocity $\omega$ in the laboratory frame is given by
\be E^{(1)}_{\rm kin}\simeq\frac{(\widetilde{m}[\vec{\omega}\times \vec{r}_3]+\widetilde{m}\delta \vec{v})^2|\Phi_{\pi,\sigma}|^2}{2},\,\, \widetilde{m} \delta \vec{v}_\theta =\frac{\nu}{r},\ee
cf. \cite{BaymChandler1983}, and we used the variable shift $\nabla\to \nabla -i\widetilde{m}\vec{v}$ in the gradient term of the energy density.
The $\omega, \Omega$ dependent contribution to the Gibbs  energy density in the rotation frame is as follows
\be \delta E^{(1)}[\omega, \Omega]\simeq \frac{\widetilde{m}^2([\vec{\omega}\times \vec{r}_3]+\delta \vec{v})^2|\Phi_{\pi,\sigma}|^2}{2}-\vec{l}^{(1)}_v\vec{\Omega},
 \ee
density of angular momentum is $\vec{l}=I_s\vec{\omega} + \vec{l}_v^{(1)}$, $I_s =\widetilde{m}^2 r^2 |\Phi_{\pi,\sigma}|^2$.

 For the array of vortices we have $\vec{l}=I_s\vec{\omega} + n_v\vec{l}_v^{\,(1)}$. Thus we obtain
 \begin{eqnarray}& {\cal E}[\omega, \Omega]\simeq
 J_s \omega^2/2 -J_s\omega\Omega -N_v L_v(\Omega-\omega)\nonumber\\&+N_v {\cal E}^{v}_{\rm kin}(\omega)\,,\label{deltaEomOm}
 \end{eqnarray}
 where
 \be{\cal E}^{v}_{\rm kin}(\omega)\simeq d_z \pi \nu^2 |\Phi_{\pi,\sigma}|^2\ln (\sqrt{\nu \Omega_{c2}/\omega})\,,\ee
  cf. Eq. (\ref{ekin1}), $J_s \simeq \pi R^4 d_z \widetilde{m}^2 |\Phi_{0\pi,\sigma}|^2/2$. We used that the logarithmic divergence is now cut at $r\sim R_L (\omega)$ given by Eq. (\ref{Rlat}) where now $\Omega$ should be repaced by $\omega$. Minimization of ${\cal E}^v_{\rm kin}(\omega)$ in $\omega$ gives $L_v =-\partial {\cal E}^{v}_{\rm kin}(\omega)/\partial \omega =d_z \pi \nu^2 |\Phi_{\pi,\sigma}|^2/(2\omega)$ that for $\omega=\Omega$ coincides with (\ref{Lvortlat}).
 Minimimizing  Eq. (\ref{deltaEomOm}) in $\omega$  we obtain $\omega[\Omega]\simeq \Omega -N_v(L_v+{\cal E}^v_{\rm kin}/\omega)/J_s$.

The difference $\omega -\Omega$ is proved to be negative showing that the lattice of vortices rotates with a bit smaller angular velocity than the vessel.
 For $\Omega \gg 1/(\widetilde{m}R^2)$  we recover Eq. (\ref{latdeltae}). For $\Omega_{c1}\ll \Omega \ll \Omega_{c2}$  we can put $\omega\simeq \Omega$, as it has been done in the paper body.
 %%%%%%%%%%%%%%

\end{document}